\documentclass[twocolumn,prx,groupedaddress]{revtex4-1}

\usepackage[utf8]{inputenc}
\usepackage[english]{babel}
\usepackage[T1]{fontenc}
\usepackage{lmodern}
\usepackage{graphicx}
\usepackage{amsmath}
\usepackage{amsfonts}
\usepackage{amssymb}
\usepackage{microtype}
\usepackage{braket}
\usepackage{caption}
\usepackage{subcaption}

\usepackage[breaklinks=true,colorlinks, citecolor=blue]{hyperref}

\newcommand{\bi}{\bibitem}
\newcommand{\cc}{\captionsetup{justification=raggedright,singlelinecheck=false}}

\newcommand{\ct}{\cite}

\newcommand{\beq}{\begin{eqnarray}}
\newcommand{\eeq}{\eeq{eqnarray}}

\begin{document}

\begin{@twocolumnfalse}

	\title{Dissipative preparation of many-body Floquet Chern  insulators}
	\author{Souvik Bandyopadhyay}
	\email{souvik@iitk.ac.in}
	\author{Amit Dutta}
	\affiliation{Department of Physics, Indian Institute of Technology, Kanpur, Kanpur 208016, India}

	\begin{abstract}
	Considering coupling to a micro-structured bath as a relaxation mechanism in a periodically driven dissipative Haldane model, we establish that the system may be tuned to a stroboscopic topological steady state at all finite temperatures. The amplitude and frequency of the periodic drive is so chosen that the Floquet Hamiltonian describing the Haldane model at stroboscopic instants of time in the unitary situation is topologically non-trivial. We establish that in the stroboscopic steady state, the system reaches a thermal state of the Floquet Hamiltonian at a controlled temperature. Further, it is  observed that even with a coupling to a quasi-local bath, remarkably  a Chern insulator can indeed be prepared in a Chern non-trivial pure steady state which is expected to exhibit a stroboscopic bulk-boundary correspondence. 
Using the non-uniqueness of the macroscopic bulk electric polarisation of a Chern insulator in its topological phase,
 we propose a generalised  Chern invariant that  reflects the topology of out-of-equilibrium many-body stroboscopic states of 
the Haldane model  even in a dissipative ambience. 
 The generalised topology of dynamical Chern insulators being dependent on  single-particle correlations, is expected to manifest in experiments probing many-body quantum observables.
		
	\end{abstract}
	\maketitle
\end{@twocolumnfalse}

\section{Introduction}
Equilibrium topological properties of quantum matter are established to be extremely robust against external local perturbations and therefore host an enormous multitude of possibilities in understanding many body quantum phases stable under experimental situations. Such topological phases are characterized by different quantized values of a topological invariant which serves as a non-local order parameter characterizing the phases which are topologically inequivalent to each other. Distinct topological phases in thermodynamically large systems are neccessarily separated by a quantum critical point (QCP) \ct{sachdev10,dutta15}, where the topology of the system becomes ill-defined. This ensures that the different topological phases of matter cannot be adiabatically connected to each other.
This has led to a large number of theoretical \ct{kitaev01,kane05,bernevig06,fu08,zhang08,sato09,sau10a,sau10b,lutchyn10,oreg10,moore10,shen12,bernevig13,haldane83,wen95,kitaev03,kitaev06,levin06} and   experimental \ct{mourik12,rokhinson12,deng12,das12,churchill13,finck13,alicea12,leijnse12,beenakker13,stanescu13} studies probing the generation and manipulation of many body topological phases of quantum systems.\\

Topological quantum systems in arbitrary dimensions are broadly characterized into having two types of topological properties. Systems exhibiting intrinsic topological order has been established to host degenerate ground states manifolds having a non-local contribution to entanglement entropy \ct{haldane83,wen95,kitaev03,kitaev06,levin06}. On the other hand, symmetry protected topological insulators (see \ct{moore10,shen12,bernevig13}, for review) and Chern insulators \ct{haldane83} host no long range topologically ordered states. However, the bulk topological non-triviality of SPTs and Chern insulators is manifested in the presence of topologically protected boundary-localised zero energy states when the bulk system is topologically non-trivial.  This bulk-boundary correspondence promises extensive applications in a plethora of emerging areas of interest such as robust quantum computation and quantum information related studies.\\

Many recent theoretical and experimental studies have started to probe the existence of such topological phases in far from equilibrium scenarios.  Dynamical engineering of  topological phases 
	in an out-of-equilibrium  state is a challenging  topic of ongoing research \ct{oka09,bermudez09,kitagawa11,lindner11,cayssol13,rudner13,patel13,thakurathi13,kundu13,rajak14,balseiro14,mitra15,gil16}
	{and so is dynamically emergent topology \ct{budich16,utso171,hu20} in context of dynamical quantum phase transitions \ct{heyl13,sharma16,heyl18} especially in topological models \ct{vajna15,utso173,dutta17,halimeh19}, both closely  connected
	to topological quantum computations \ct{kitaev16,pachos17}.
	The success of such dynamical preparation depends not only on the dynamical generation of a topological Hamiltonian but also
on preparing the system in a topologically non-trivial dynamical state. The question whether the out-equilibrium state of a quantum many body system can be a characterised
by an integer-quantised topological index  and whether there exist a non-equilibrium bulk-boundary correspondence has not yet been fully understood \ct{foster13,foster14,rigol15,cooper15,utso17,sougato18,ginley18,souvik90,souvik191,pastori20}.  The dynamical topological invariant has been recently studied in out-of-equilibrium one dimensional (1D) topological system\ct{foster13,foster14,ginley18,souvik90,souvik191}.  Interestingly,  in Ref. \ct{souvik191}
 using  a periodic driving scheme with a linearly ramped amplitude, 
 a stroboscopic  "out of equilibrium" bulk-boundary correspondence has been established for 1D SSH and extended SSH models. 
 It has thus been established that it is indeed possible to dynamically construct a topological non-equilibrium state having gapless boundary localised excitations. Further, the inter-relation between different probes such as the entanglement entropy in the occurrence of dynamical quantum phase transitions \ct{heyl13,sharma16,heyl18} and the role of interactions have been  explored in Ref.~[\onlinecite{pastori20}].

Interestingly, for two-dimensional (2D) Chern insulating systems obeying periodic boundary  conditions,  it has been argued \ct{rigol15} that it is not possible to change the initial  topology of the  model through a smooth unitary transformation: this is a consequence of the  temporal invariance of the dynamical bulk Chern number, extracted from the time evolved state of the system under such driving. 
However, following a quenching  from the topological phase to a trivial phase, the edge current is found to vanish asymptotically, implying that the edge current eventually thermalizes to a value corresponding to the topology of the post-quench Hamiltonian \ct{cooper15,utso17,sougato18}; thereby implying the absence of an out-of-equilibrium bulk-boundary correspondence. 
Incidentally, generalising  the non-uniqueness of the bulk polorisation of  a CI in its topological phase at equilibrium,
a dynamical generalised Chern invariant has been proposed.  This quantity indeed captures the  out-of-equilibrium topology  of the model under an  adiabatic temporal evolution within a unitary driving protocol \ct{souvik201}.\\

 In parallel, there has been a plethora of studies  concerning the dissipative preparation of topological states in open systems \ct{diehl08,diehl11,bardyn13,budich15,goldman16,diehl18}, the fate of equilibrium topology due to the coupling to external baths \ct{kraus12,carmele15,souvik20} and mixed state emergent topology \ct{utso172,budich17,souvik18} and propagation of correlations in
 out-of-equilibrium open quantum systems \ct{maity20,alba20}. In particular, it was argued in Ref.~[\onlinecite{budich15}] that it is indeed  possible  to define a  Chern number using the time evolved density matrix  which may change  dynamically  under a temporal evolution. Furthermore,  it has also been established that the Chern invariant can also assume topologically quantised values in the asymptotic steady state provided the steady state is mixed. Despite the {possible} dynamical variation of the Chern invariant, a generic validation of the conventional bulk-boundary for mixed states is still lacking. Therefore, as a consequence,  in far from equilibrium situations, the familiar notion of the bulk-boundary correspondence apparently breaks down.

Thus, the emergence of a topological bulk-boundary correspondence in a non-equilibrium or an asymptotic steady state is only possible if the corresponding state is pure.  However, it has been reported recently, that a Chern non-trivial {\it pure} steady state cannot be the asymptotic steady state of a dissipative system in the Lindblad master equation approach as long as the jump operators have finite length scale of action  
 \ct{bardyn13,goldstein19}. This makes the study of the out-of-equilibrium bulk boundary correspondence for a generic dynamics far more complicated within a Lindblad framework. Whether the locality constraint on the action of the bath remains valid in a driven dissipative system, is an exciting question to address.\\

In this work, we propose an alternative path to surmount the hurdles mentioned above by  investigating the topological properties of Chern insulators when the system is driven out-of equilibrium in terms of many-body observables quantities. Questions we address are the following: (i) Is it at all possible for out of equilibrium Chern insulators to exhibit topological properties or are all equilibrium topological properties washed away far from equilibrium? (ii) Is it possible to reach both mixed and pure steady topological states in driven dissipative Chern insulators in the presence of a bath  with observable many-body topological properties? \\

We establish that a periodically driven 2D  Chern insulator, {namely the Haldane model of graphene}, may be dynamically prepared into a {stroboscopic steady state}  arbitrarily close to a many-body Chern insulating Floquet topological phase in the presence of a fermionic bath \ct{iwahori16} which acts quasi-locally on the model. In doing so, we construct a micro-structured reservoir which may be coupled to the graphene sheet  as a substrate {which absorbs the excess energy transferred  to the system due to continuous external pumping.}
{The frequency and amplitude of the periodic driving is chosen in such a 
manner that the Floquet Hamiltonian describing the closed  Haldane Hamiltonian is topological, thereby ensuring  that the system may indeed thermalize into a topologically non-trivial thermal steady state in the dissipative ambience.}
We observe the topological invariant through the two-point equal time correlations by constructing the many-body macroscopic bulk electric polarisation of the system. Such an approach has been proved to be effective in capturing the topology of mixed Gaussian states of 1D systems \ct{diehl18}. Although it was suggested that generalising the same to the Chern insulator would be interesting, to the best of
our knowledge, the present work resolves the issue for the first time. We achieve this  by avoiding the Lindbladian approach and using this specially designed bath required to  facilitate the preparation of a Gaussian stroboscopic steady state. \\

{What complicates the scenario is that unlike} 1D topological systems such as the SSH model, the bulk electric polarisation is not a topologically quantised quantity in itself {in 2D system. Nevertheless,} the electric polarisation in Chern insulators is known to exhibit unphysical non-uniqueness properties when the system is in a Chern non-trivial phase \ct{vanderbilt09}. More precisely, the bulk electric polarisation in a Chern insulator is uniquely defined only for a particularly specified Brillouin zone (BZ). A universal {adiabatic} translation in each momentum vector within the BZ, shifts the value of the electric polarisation by a factor proportional to the Chern invariant. This approach has 
already been successfully employed to define a dynamical Chern invariant for a closed (pure state) out-of-equilibrium Chern insulator  within a unitary protocol \ct{souvik201}.\\

We exploit this property of a Chern insulator to a periodically driven dissipative Haldane model {starting from the topologically trivial state
of the bare Haldane model}. We establish that although the bath acts locally on the lattice, the asymptotic steady state can indeed be topologically non-trivial even for both pure and mixed steady states. We also probe the existence of the bulk-boundary correspondence in the steady state when it is pure. The defined Chern number having a many-body nature, is also expected to exhibit many-body topological properties of the system. Following this, we argue that even if the steady state is a mixed topological state, its topological non-triviality is expected to manifest in interferometric setups where the system interacts with certain cavity modes of electromagnetic radiation.\\

The paper is organised in the following manner. 
In Sec.~\ref{sec:hamiltonian} we start with the description of the model studied and the bath chosen to include a relaxation mechanism. The periodic driving protocol and the corresponding Floquet Hamiltonian is also introduced in this section.  \\In Sec.~\ref{sec:eom}, we explicitly write down the Heisenberg equations of motion of the system and bath degrees of freedom and then solve them simultaneously in the asymptotic limit under weak coupling and high frequency approximation in Sec.~\ref{sec:rot}. In Sec.~\ref{sec:pop}, we proceed to characterize the steady state of the reduced system by integrating out the bath degrees of freedom and establish that the stroboscopically observed system indeed thermalizes into a Floquet Gibbs ensemble, so that the density matrix assumes a Gaussian form.\\ In the next section i.e., in Sec.~\ref{sec:epol} we define the macroscopic polarisation for the steady state stroboscopic system using the two point correlation functions and noting it's non-uniqueness in a non-trivial topological phase, we define a many body Chern number in Sec.~\ref{sec:chern}. We establish that the stroboscopic steady state of the reduced system exhibits topological non-triviality through a non-zero Chern number. \\ In Sec.~\ref{sec:current}, we proceed to calculate the system-bath particle current and establish that the mean number of particles within the system approaches a steady value in the stroboscopic steady state which justifies the Gaussian nature of the steady state reduced density matrix of the system. 
{In this section, we also discuss the existence of a topological bulk-boundary  correspondence when the stroboscopic steady
state is pure.} Concluding comments and possibilities of experimental detections are discussed in Sec. \ref{sec_conclusion}. {We have
further included three appendices: appendix \ref{appendix:C} contains a review of the Haldane model of graphene while  \ref{Sec:solution} provides involved  calculational detail. Finally, in appendix \ref{sec:bulk_pol_ap}, we discuss the detail of the many-body macroscopic electrical polarisation.
}


\section{The driven dissipative bulk system}\label{sec:hamiltonian}
We start with a quadratic fermionic 2D  having a sub-lattice structure (particularly the Haldane model of {graphene} )
subjected to a generic temporal drive generated by the Hamiltonian  $H_S(t)$. The system is assumed to be coupled to a free fermionic quasi-local bath $H_R$ (the {\it reservoir}) through a bilinear coupling $H_I$. The time evolution of the complete system is then described by the Hamiltonian,
\begin{equation}
H(t)=H_S(t)+H_R+H_I,
\end{equation} 
\begin{figure}

\includegraphics[width=\columnwidth]{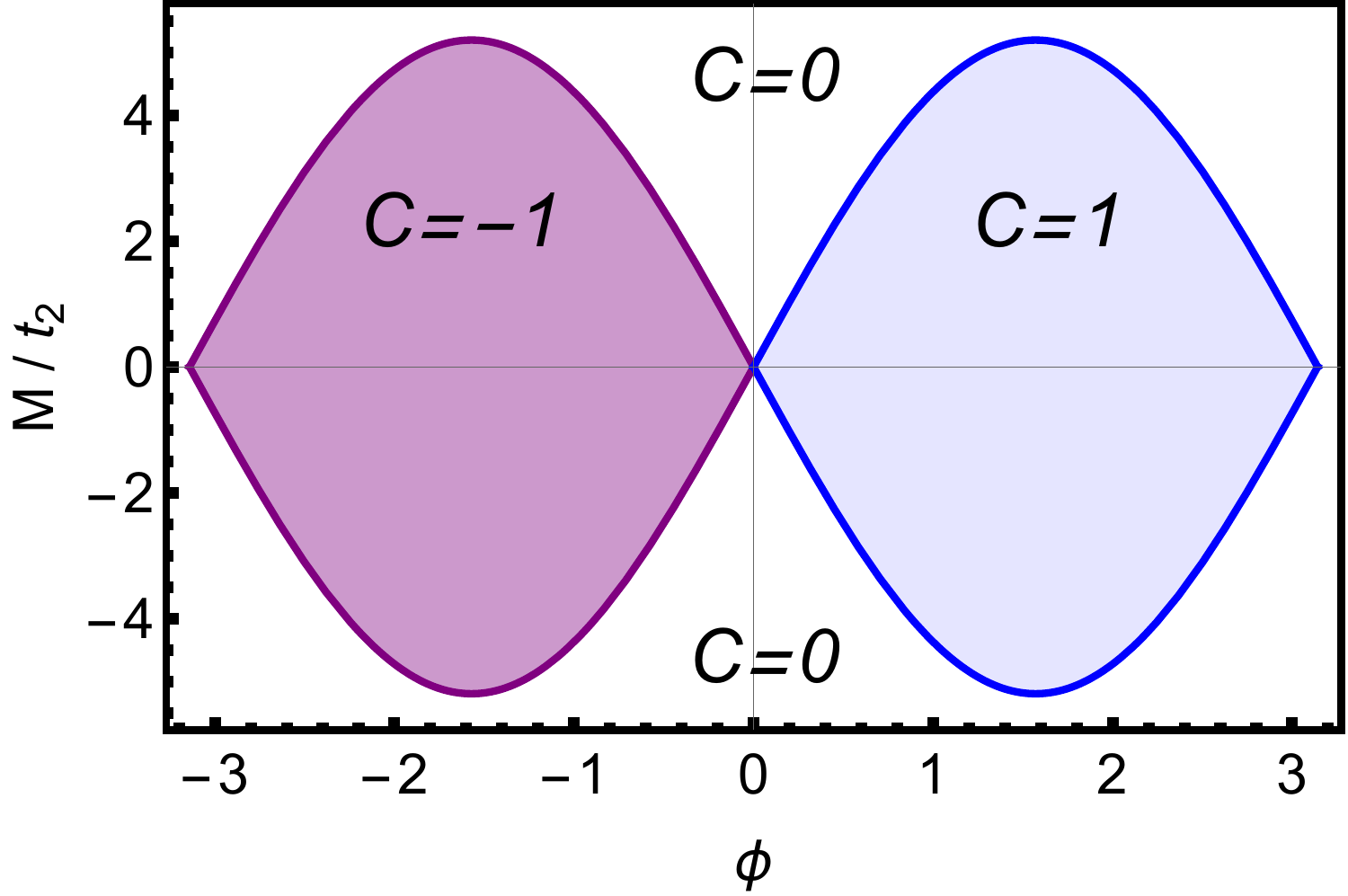}

\caption{ (Color online) The topological phase diagram of the Haldane model {with the nearest neighbour hopping $t_1=1$}. The distinct topological phases are sepatrated by critical lines on which the parameter values are such that the system becomes gapless. The parameter regions showing non-zero values of the Chern number $C$ are topologically non-trivial.}

\label{fig:1}

\end{figure}

\begin{figure*}
	\centering
	\begin{subfigure}{0.4\textwidth}
		\centering
		\includegraphics[width=\columnwidth]{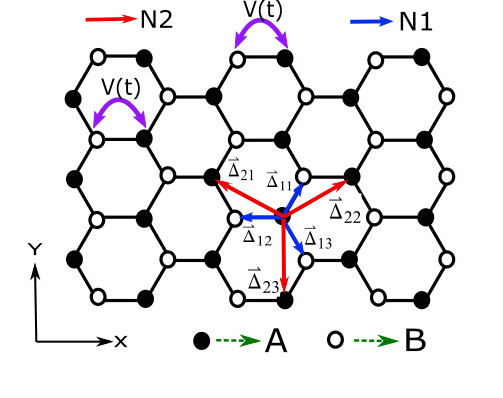}
		\caption{}
		\label{fig:2a} 
	\end{subfigure}
	\begin{subfigure}{0.4\textwidth}
		\centering
		\includegraphics[width=\columnwidth]{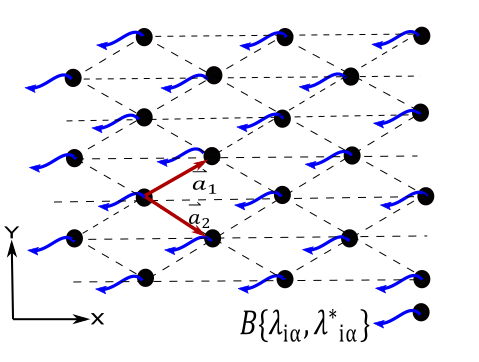}
		\caption{}	\label{fig:2b}
	\end{subfigure}\quad\quad\quad\quad

	\cc
	\caption{ (Color online) (a) The honeycomb lattice of the Haldane model showing the sublattice structure and the action of the time-periodic drive as in Eq.~\eqref{eq:driven H} acting on nearest neighbour hopping. The periodic potential $\hat{V}(t)$ induces a complex anisotropic nearest-neighbour hopping which is periodic in time.({See Appendix.~\ref{appendix:C}})
	(b) The coupling to the reservoir and the system-reservoir interaction. Each basis is independently coupled to a free-fermionic bath ${\mathcal B}$ with different coupling strengths for the $A$ and $B$ sites but uniform coupling for all Bravias lattice sites. The bath degrees of freedom coupled to different sites on the Bravias lattice are necessarily independent to preserve the sublattice structure in the steady state.
	}
	
\end{figure*}
\noindent such that,
\begin{equation}
\begin{split}
H_S(t)=\sum_{\alpha,\beta,n,m} H_{\alpha,\beta,n,m}(t) a_{m,\alpha}^{\dagger}a_{n,\beta}+h.c.,\\
H_R=\sum_{i,m}\epsilon_{i}A_{m,i}^{\dagger}A_{m,i}~~\text{and,}\\
H_I=\sum_{i,n,\alpha}\lambda_{i,\alpha} A_{n,i}^{\dagger}a_{n,\alpha}+h.c.,
\end{split}
\end{equation}
where $a_{n,\alpha}$ and $A_{i}$ satisfies fermionic anti-commutation relations independently. The indices $n$ and $\alpha$ on $a_{n,\alpha},A_{n,\alpha}$
denote the sublattice and intra-sublattice index respectively, i.e., $\alpha\in\{A,B\}$ and $n\equiv\{n_1,n_2\}$ is the position of a site in the Bravias lattice having a two-point basis. Depending on the parameters the Haldane model hosts a non-trivial topological phases (Fig.~\ref{fig:1}). We focus on a situation in which the dissipative Haldane model is driven periodically in time (Fig.~\ref{fig:2a}-\ref{fig:2b}), i.e.,
\begin{equation}
H_{\alpha,\beta,n,m}(t)=H_{\alpha,\beta,n,m}^0+V_{\alpha,\beta,n,m}(t),
\end{equation} 
where $H^0(M,t_1,t_2,\phi)$ is the bare Haldane model. 

The bare Hamiltonian for the Haldane model is obtained by breaking the time reversal and sublattice of graphene,
\begin{equation}
\begin{split}
H_{\alpha,\beta,n,m}^0=-t_1\sum\limits_{\left<m\alpha,n\beta\right>}a_{m,\alpha}^{\dagger}a_{n,\beta}+M\sum\limits_{n} a_{n,A}^{\dagger}a_{n,A}\\
-M\sum\limits_{n} a_{n,B}^{\dagger}a_{n,B}-\sum\limits_{\left<<m\alpha,n\alpha\right>>}t_2e^{i\phi}a_{m,\alpha}^{\dagger}a_{n,\alpha}+h.c.,
\end{split}
\end{equation}	
where the real hopping $t$ comprises the bare graphene Hamiltonian. The diagonal staggered mass (Semenoff mass) $M$ explicitly breaks the sublattice symmetry of the model. Further the complex next nearest neighbour hopping term $t_2$, is applied such that the time reversal symmetry is broken in the next nearest neighbour hopping while the net flux through each plaquette remains zero.
The Haldane model is known to exhibit non-trivial Chern topology when its ground state is completely filled depending on the parameters $M$, $t_1$, $t_2$ and $\phi$ characterised by the Chern number $C$.
{Interestingly, the Haldane model with explicitly broken time reversal symmetry is known to host topologically non-trivial phases for certain parameter regions having a non-zero Chern number (see Fig.~\ref{fig:1}).\\
	The Chern invariant is integer quantized as long as the Hamiltonian $H^k$ does not approach a quantum critical point where the Chern number becomes ill-defined. Different integer values of the Chern number characterize distinct topological phases separated by QCPs (see Fig.~\ref{fig:1}). 
}

{To retain the discrete translation symmetry of the system { in the presence of coupling to the bath}, we assume mutually decoupled local baths ${\mathcal B_i}$ which individually couple to each} 
{Bravias lattice point independently. This ensures that no inter-sublattice hoppings are introduced due to coupling to the bath.
Each point on the Bravias lattice can be referenced in terms of the Bravias lattice vectors, i.e.,
\begin{equation}
\vec{a}=n_1\vec{a}_1+n_2\vec{a}_2,
\end{equation}
where the vectors $\vec{a}_1$ and $\vec{a}_2$ span the Bravias lattice and $n_1,n_2$ are integers.
 Invoking the discrete translational invariance of the Hamiltonian one can employ discrete Fourier transform to decouple the Hamiltonian $H(t)$ in momentum space. The reciprocal space is spanned by the reciprocal lattice vectors $\vec{b}_1$ and $\vec{b}_2$, i.e. every reciprocal lattice point can be represented as,
 \begin{equation}
 \vec{b}=k_1\vec{b}_1+k_2\vec{b}_2,
 \end{equation}
 where, $k_1$, $k_2\in[0,1]$ ({See Appendix.~\ref{appendix:C} for detail}}).

{Preparing  the model initially at time $t = 0$  in the topologically trivial phase having Chern number $\mathcal{C}=0$, we  subject it to a periodic driving  $V(t+T)=V(t)$ with driving of frequency $\omega=2\pi/T$ such that it solely acts on the nearest-neighbour hopping amplitudes (see Fig.~\ref{fig:2a}).}
{The corresponding single-particle Hamiltonian therefore decoupled for each momenta mode,
\begin{equation}\label{Hk}
\begin{split}
H^k_{full}(t)=\bigoplus_{k}\sum_{\alpha,\beta} H_{\alpha,\beta}^k(t)a_{\alpha}^{k\dagger}a_{\beta}^k+\sum_{i,k}\epsilon_i^kA_{i}^{k\dagger}A_{i}^k\\
+\sum_{i,\alpha}\lambda_{i,\alpha}A_{i}^{k\dagger}a_{\alpha}^k+h.c.,
\end{split}
\end{equation} 
where $k$ denotes the ordered pair $\left(k_1,k_2\right)$ and
$H^k(t)$ is the quasi-momentum resolved bare Haldane Hamiltonian subjected to the periodic perturbation given as,
\begin{equation}\label{eq:driven H}
H^k(t)=H^0(k)+V(t)=H^0(k)+V_0\left[\sigma_x\cos{\omega t}+\sigma_y\sin{\omega t}\right],
\end{equation}
where $H^0(k)$ is the bare Haldane Hamiltonian in momentum space can be written in the basis $\ket{k,A}$ and $\ket{k,B}$ as,
\begin{equation}
H^0(k)=\vec{h}(k).\vec{\sigma} =h_x(k) \sigma_x+h_y(k) \sigma_y+h_z(k) \sigma_z,
\end{equation}
such that,
\begin{equation}\label{eq:bloch_ham}
\begin{split}
h_x(k)=-t_1\sum\limits_{i=1}^{3}\cos{\left(\vec{k}.\vec{\Delta}_{1i}\right)},\\
h_y(k)=-t_1\sum\limits_{i=1}^{3}\sin{\left(\vec{k}.\vec{\Delta}_{1i}\right)},\\
h_z(k)=M-t_2\sin{\phi}\sum\limits_{i=1}^{3}\sin{\left(\vec{k}.\vec{\Delta}_{2i}\right)},
\end{split}
\end{equation}
$\vec{\Delta}_{1i}$ and $\vec{\Delta}_{2i}$ are the nearest neighbour and next nearest neighbour lattice vectors respectively.}\\

{{Let us recall the purely unitary evolution of the Haldane model dictated by the Hamiltonian in Eq.~\eqref{eq:driven H},  starting from a non-topological  ground state. The
corresponding Floquet Hamiltonian $H_F(k)$ generating
the unitary stroboscopic evolution for $t>0$, in the high frequency limit of driving assumes the form \ct{kitagawa11},}
\begin{equation}
H^F(k)\simeq H^0(k)+\frac{1}{\omega}\left[H_{-1}(k),H_{+1}(k)\right]=H^0(k)-\frac{V_0^2}{2\omega}\sigma_z,
\end{equation}
\begin{figure}
\begin{center}
\includegraphics[width=0.48\textwidth,height=0.65\columnwidth]{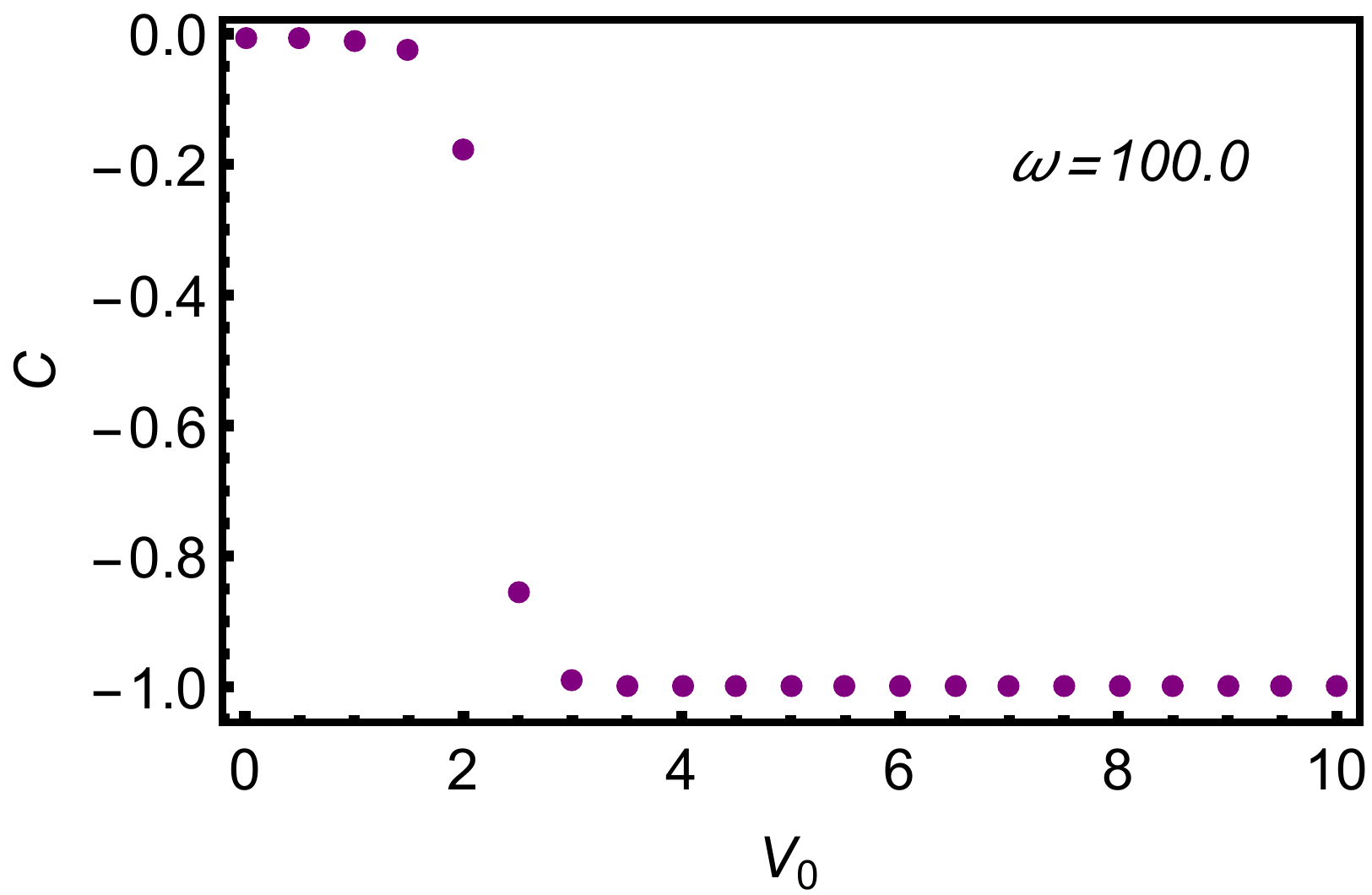}
\end{center}

\caption{(Color online) The variation of the Floquet Chern number $C$ with the driving amplitude under a high frequency drive (see Eq.~\eqref{eq:driven H}) is shown to assume a topological value (in this case, $C=-1$) with a certain driving amplitude. The driving frequency $\omega$ is chosen to be $\omega=100.0$. The Haldane Hamiltonian is chosen with $t_1=1.0$, $t_2=0.5$, $M=3\sqrt{3}t_2+0.05$ and $\phi=-\frac{\pi}{2}$ for a $200\times 200$ lattice with a periodic boundary condition.}
\label{fig:3}  
\end{figure}

\noindent to the leading order of $V_0^2/\omega$.
Evidently, the chiral symmetry breaking Semenoff mass of the bare Hamiltonian
$H^0(k)$ (appearing in $h_z(k)$) gets renormalised in the Floquet Hamiltonian.
However, in this work the amplitude $V_0$ is so chosen to
ensure that the Floquet Hamiltonian $H^F(k)$ lies in the
topological sector of the Haldane phase diagram in contrary to the initial Hamiltonian $H^0(k)$ (see Fig.~\ref{fig:3}). The Chern number $C$ is evaluated as (see Appendix.~\ref{appendix:C}),
\begin{equation}
C=\frac{1}{\left(2\pi\right)^2}\int_{BZ}dk_1dk_2\mathcal{F}_{12}(\ket{g_k}),
\end{equation}
where $\mathcal{F}_{12}(\ket{g_k})$ is the $U(1)$ curvature of the Floquet eigenstate $\ket{g_k}$ having lowest quasi-energy \ct{kitagawa11}.
{The topological nature
of the Floquet Hamiltonian is crucial in determining the topological nature of the stroboscopic steady state reached in 
the presence of the bath. }

\section{Equations of motion}\label{sec:eom}

In this section, we shall consider the temporal evolution
of the Haldane model under the periodic driving in the
presence of the bath considering the Hamiltonians Eq.~\eqref{Hk}
and {the periodic driving} as in Eq.~\eqref{eq:driven H}.
{As already discussed in the introduction, within the Lindbladian approach  arriving at an asymptotic topological steady state is complicated  as long as the jump operators have
finite length of action \ct{bardyn13,goldstein19}.
Avoiding a Lindbladian approach}, we therefore make resort to the Heisenberg picture and determine the
dynamical equations of motion of the operators $a^k_{\alpha}$ and $A^k_{i}$ {which assume the form:}

\begin{equation}\label{eq:eom}
\begin{split}
i\frac{dA^k_i}{dt}=\epsilon_i^kA_i^k(t)+\sum_{\alpha}\lambda_{i,\alpha}a_{\alpha}^k(t)\\
i\frac{da_{\alpha}^k}{dt}=\sum_{\beta} H_{\alpha,\beta}^k(t)a_{\beta}^k(t)+\sum_{\mu}\lambda^{*}_{\mu,\alpha}A_{\mu}^k(t).
\end{split}
\end{equation}
{ Eqs.~\eqref{eq:eom} encode} the dynamics of the system and as well as the  bath degrees of freedom for the driven composite system. {We shall now  simultaneously solve the above set and eliminate the bath degrees of freedom to obtain the dynamics of the system operators.} 

The explicit time dependence of the system  Hamiltonian $H_s$ can be completely eliminated using a time periodic unitary transformation (see Appendix~\ref{Sec:solution}). The resulting unitary part of the dynamics is then governed by an effective Hamiltonian $H^{\rm eff}_k$ with no explicit time variation.}
{After eliminating the reservoir degrees of freedom and recasting the dynamics in terms of the new operators  $f_b^k(t)$} which are eigen-operators of $H^{\rm eff}_k$ and the index $b$ here signifies the different Floquet bands,
  
 \begin{eqnarray}\label{eq:flo_mode_dyn}
 i\partial_t f_b^k &=& E^k_b f_b^k(t)\nonumber\\
&-i&\sum_{m,n,b^{\prime}}e^{i(n-m)\omega t}\int_{0}^t\tilde{\Pi}^{k,nm}_{b,b^{\prime}}f_{b^{\prime}}^k(t-t^\prime)e^{im\omega t^\prime}dt^\prime\nonumber\\
&+&i\sum_{b^{\prime}}Y^{k\dagger}_{b,b^{\prime}}\zeta_{b^{\prime}}^k(t),
\label{eq:Floquet_eqn}
\end{eqnarray}
where,
$E_{b}^k$ are the eigenvalues of the time-independent effective Hamiltonian $H^{\rm eff}_k$ and,
\begin{equation}\label{eq:Pi}
\begin{split}
\tilde{\Pi}^{k,nm}_{b,b^{\prime}}=\left[Y^{k(n)\dagger}\Pi^k(t)Y^{k(m)}\right]_{b,b^{\prime}},\\
Y^{k(m)}(\omega)=\frac{1}{T}\int_{0}^{T}Y^k(t)e^{im\omega t}dt.
\end{split}
\end{equation}

The dynamics of the Floquet operators are generated by essentially two kinds of processes, the coherent unitary driving and scattering due to coupling to the bath. The first term on the right hand side of Eq.~\eqref{eq:flo_mode_dyn}, signifies the unitary diagonal evolution of the Floquet modes while the second and third terms entail the dissipative processes. The quantity $\tilde{\Pi}^{k,nm}_{b,b^{\prime}}$ as defined in Eq.~\eqref{eq:Pi} is the scattering amplitude collating all scattering processes between different photon sectors and Floquet bands of the drive while $\zeta_{\beta}^k(t)$ is the noise kernel resulting from interaction of the system with the dissipative reservoir.

The dynamics generated by Eq.~\eqref{eq:Floquet_eqn} can be further simplified under assumptions of weak coupling and high frequency. In next two sections, we proceed to simplify the dynamical equation and extract from it relevant information about the asymptotic steady state.

\section{Weak coupling and rotating wave approximation}\label{sec:rot}

{Having set the dynamical equation of motion of the system variables}, we now proceed to identify the relevant time scales in the problem and quantitatively recognize the asymptotic times in which the steady state solution is expected. {Although generically  Eq.~\eqref{eq:Floquet_eqn}  is valid for an arbitrary  coupling strength between the system and the bath, henceforth we shall  employ a weak coupling approximation.} We elaborate on the approximations which simplify Eq.~\eqref{eq:Floquet_eqn} at large times, 
\begin{itemize}
\item We assume that the collective coupling to all modes of the bath is insufficient to induce direct transitions in asymptotic time between different energy states (if they differ in energy) of the effective Hamiltonian $H^{\rm eff}_k$. The coupling to the fermionic dissipator is also assumed to be insufficient to induce direct transitions between different photon sectors in the steady state, i.e. we choose to observe the system much later to a time scale $t_s^{(1)}$ such that,
\begin{equation}\label{eq:approx_1}
\frac{\Lambda^k_{b,b^{\prime}}\left(t_s^{(1)}\right)}{E_{b}^k-E_{b^{\prime}}^k+n\omega}\ll 1 ~~\forall~b\neq b^{\prime},~\mu~\text{and}~n\in{\mathrm Z},\\
\end{equation}
where $\Lambda^k_{b,b^{\prime}}(t_s^{(1)})$ includes the total scattering between the bands $b$ and $b^{\prime}$ in time $t_s^{(1)}$.
The time-scale $t_s^{(1)}$ resembles the approach to the dissipative steady state under the action of the bath \ct{iwahori16}. The approximation in Eq.~\eqref{eq:approx_1}, intuitively signifies that after time $t_s^{(1)}$, the scattering between two Floquet bands proportional to the quantity, $\lambda^*_{\mu,\alpha}\lambda_{\mu,\beta}t_s^{(1)}$ is much weaker for two non-degenerate Floquet bands of different quasi-energies. 

\item Moreover, at asymptotic times $\left(t>t_s^{(2)}~\text{such that}~\omega t_s^{(2)}\gg 1\right)$ for a high frequency drive, the second sum on the right of the equality in  Eq.~\eqref{eq:Floquet_eqn} oscillates rapidly for $n\neq m$, hence one may neglect its contribution to the equation except for $n=m$. {That is, the time-scale $t^{(2)}_s$ signifies the time at which the off-diagonal elements of the Floquet Hamiltonian decohere and all photon absorption-emission processes are suppressed.}
\end{itemize}

Under these approximations, if one observes the system at large times such that $\left(t\sim O\left(t_s^{(1)}\right)~~\text{and}~~t\gtrsim t_s^{(2)}\right)$, only the diagonal part of the matrix $\tilde{\Pi}^{k,nm}_{\alpha,\beta}$ contributes predominantly and the dynamical equation at asymptotically large times reduces to,
\begin{equation}\label{eq:diagonal Floquet}
\begin{split}
i\partial_t f_b^k=E^k_b f_b^k(t)-i\sum_{n}\int_{0}^t\tilde{\Pi}^{k,nn}_{bb}f_{b}^k(t-t^\prime)e^{in\omega t^\prime}dt^\prime\\
+i\sum_{b^{\prime}}Y^{k\dagger}_{b,b^{\prime}}\zeta_{b^{\prime}}^k(t),
\end{split}
\end{equation}
to include a diagonal $\tilde{\Pi}^{k,nn}_{b,b^{\prime}}$ which is a result of only virtual transitions and amounts to the self energy corrections due to coupling to the reservoir and time-periodic driving. Under these approximations, we proceed to construct the steady state solution of the dynamical system in Eq.~\eqref{eq:diagonal Floquet}.

\section{Bath assisted occupation of topological Floquet states}\label{sec:pop}

In this section, we show that under the assumptions of weak system-bath coupling and high frequency driving, the system reaches a Gaussian steady state when observed stroboscopically. The stroboscopic steady state being Gaussian, it is possible to exactly evaluate the two-point fermionic correlations analytically without resorting to perturbative approaches. As we shall demonstrate, the validation of Wick's theorem in the stroboscopic steady state further enables one to calculate all many-particle correlations and many-body observables exactly at asymptotic times. Moreover, we argue that since the stroboscopic steady state is a thermal state, its temperature and hence purity can be completely controlled by tuning the temperature of the bath.

Despite the continuous external driving, the system
 thermalizes to a finite temperature Gibbs' state. This can be intuitively understood as the system is allowed to discard absorbed energy from the drive into the bath until a steady equilibrium is reached. Solving the Eq.~\eqref{eq:diagonal Floquet} for the asymptotic effective Hamiltonian modes under the approximations described in Sec.~\ref{sec:rot}, we explicitly compute the asymptotic occupation of the effective modes $f_{b}^k(t)$ (See Appendix~\ref{Sec:solution} for detail).

We further assume that the free fermionic bath remains in equilibrium at all times with its energies distributed according to a Fermi-Dirac distribution at a temperature $T$ and chemical potential $\mu$, i.e., 
\begin{equation}
\left<A^{k\dagger}_{\alpha}A^k_{\alpha}\right>=f_{FD}(\epsilon^k_{\alpha},\mu=0),
\end{equation}
where we set the chemical potential of the bath $\mu$ equal to zero, i.e., within the gap of the effective Hamiltonian.
Equivalently, the reservoir has a thermal density matrix is assumed to be a Gibbs state,
\begin{equation}\label{eq:bath_thermal}
\rho_R=\bigotimes_k\frac{ e^{-\beta\sum\limits_{\alpha}\epsilon^k_{\alpha}A^{k\dagger}_{\alpha}A^k_{\alpha}}}{{\cal Z}_k},
\end{equation}
where ${\cal Z}_k$ is the normalization factor.


Solving Eq.~\eqref{eq:diagonal Floquet} to compute the fermionic occupation of the effective Hamiltonian bands and simplifying, one finally obtains,
\begin{equation}\label{eq:wei}
\left<f_{b}^{k\dagger}f_{b}^k\right>=\frac{\sum_n W^{knn}_{b b}(E^{k(n)}_{b})f_{FD}(E^{k(n)}_{b})}{\sum_n W^{knn}_{b b}(E^{k(n)}_{b})},
\end{equation}
which is a weighted average of $f_{FD}(E^k_{b},\mu)$ over all the photon sectors. The weights (see Appendix~\ref{Sec:solution}) quantify the occupation of each photon sector to the asymptotic population of the Floquet bands. As a function of energy, the weights also reflect the energy dependence contribution of the higher photon sectors in the steady occupation of the effective Hamiltonian bands.
This suggests that even if the temperature of the bath is near absolute zero, the occupation of a Floquet band have significant contribution from all the photon sectors. However, in the high frequency limit the contribution of the higher photon sectors decay,
$$W^{k(nn)}\sim{\it O}\left(\left(\frac{A^2}{\omega}\right)^{2n}\right).$$
Moreover, if an energy cutoff $\Omega_c$ is introduced in the dissipative coupling  {such that,}
the contribution of the higher photon sectors reduce significantly in Eq.~\eqref{eq:wei},i,e, if
\begin{equation}
\left|\frac{W^k(E_1)}{W^k(E_2)}\right|\rightarrow 0,
\end{equation}
for all $|E_1|\gg \Omega_c$ and $|E_2|\ll \Omega_c$, the only Floquet sectors that contribute to the sum in Eq.~\eqref{eq:wei} are such that $|E^{k(n)}_{\alpha}|<\Omega_c$. {Such a cut-off in the system bath coupling 
	
\begin{figure}
	
	\begin{center}
		\includegraphics[width=0.48\textwidth,height=0.65\columnwidth]{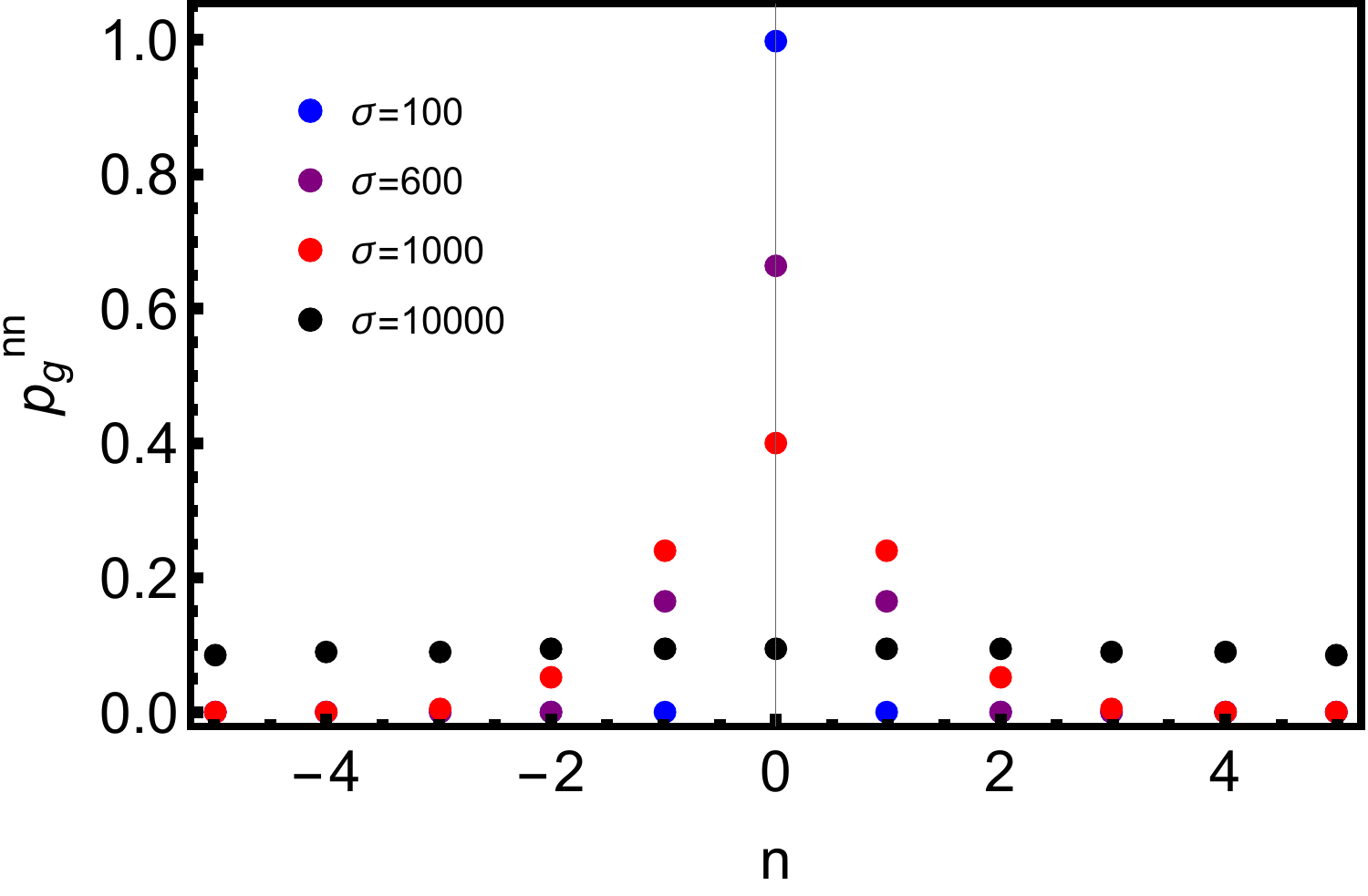}
	\end{center}

	\caption{ (Color online) The differential contribution of different photon sectors (Eq.~\eqref{eq:diff_pol}) to the steady state occupation of the Floquet bands. The system bath coupling strength is chosen from a Gaussian distribution (Eq.~\eqref{eq:cutoff}) of coupling energy having a mean and standard deviation of $m=2$ and $\sigma$ respectively. The bare Haldane model is chosen such that $M=3\sqrt{3}t_2+0.025$, $t_2=0.1$, $t_1=0.5$, $\phi=\pi/2$ and having $60\times60$ lattice points under periodic boundary conditions is subjected to a periodic driving protocol as in Eq.~\eqref{eq:driven H} with $V_0=10.0$ and $\omega=1000$. The stroboscopic steady state is taken to be a Gibbs state as in Eq.~\eqref{eq: density_matrix} having an inverse-temperature of $\beta=10$. The figure shows that that using a energy cut-off in the system bath coupling and under a low bath temperature it is indeed possible to populate a single band in a selective Floquet sector.}
	\label{fig:diff_pop}

\end{figure}

ensures that the energy window of interaction between the system and reservoir is finite.}
 If the driving frequency is high enough $(\omega\gg\Omega_c)$, solely the zero photon sector contributes significantly in Eq.~\eqref{eq:wei} and the system thermalizes into an effective Fermi-Dirac distribution.
We recall that the single-particle correlators are diagonal in the band space, i.e,
\begin{equation}
\left<f_{b}^{k\dagger}f^{k^{\prime}}_{b^{\prime}}\right>\simeq\delta_{bb^{\prime}}^{k k^{\prime}}f_{FD}(E^{k(n=0)}_b).
\end{equation}

Therefore in the presence of an energy cutoff to the system bath coupling, the zero photon sector of the Floquet Hamiltonian is predominantly occupied if further the temperature and the chemical potential of the fermionic bath is chosen to be zero.\\

{Since, in the high frequency limit, the contribution of the higher Floquet sectors are anyway suppressed, we demonstrate this selective occupation of the Floquet bands in a Gibb's steady state simply with the undressed scattering amplitudes $W^k(\Omega)$. We show that imposing a cut-off in the system-bath coupling strength, it is indeed possible to selectively populate a single photon sector. The coupling to the bath is assumed to have a Gaussian cut-off in Fig.~\ref{fig:diff_pop}. The coupling strength of the reservoir with the system is assumed to depend on the energy through a normal distribution having mean $m$ and standard distribution $\sigma$,
\begin{equation}\label{eq:cutoff}
W_{bb^{\prime}}^k(\Omega)=\delta_{bb^{\prime}} e^{-\frac{\left(\Omega-m\right)^2}{2\sigma^2}}.
\end{equation} 
For this distribution of coupling constants, we evaluate the total contribution of a distinct Floquet sector to the ground state occupation,
\begin{equation}\label{eq:diff_pol}
p_g^{nn}=\frac{1}{L^2}\sum_{k}{\frac{W^{k}_{gg}(E^{k(n)}_{g})f_{FD}(E^{k(n)}_{g})}{\sum\limits_{\nu} W^{k}_{gg}(E^{k(\nu)}_{g})}},
\end{equation}
for a $L\times L$ lattice and where $E^{k(\nu)}_{g}$ denotes the quasi-energy of the lower energy Floquet band.
The Fermi distribution $f_{FD}$ is taken to be at a low temperature steady state in equilibrium with the Floquet Hamiltonian. As observed in Fig.~\ref{fig:diff_pop}, the contribution of higher photon sectors drops significantly as the energy bandwidth of the reservoir coupling (proportional to the standard deviation $\sigma$) is made much smaller than the driving frequency.}

 We also note that the correlations among the effective fermionic operators assume a diagonal stationary form at asymptotically large times. The total number of particles in the system assumes a constant value in the stroboscopic steady state, i.e. $\partial_t\braket{N}_{\rho(\infty)}=0$ (see Sec.~\ref{sec:current}). 
Further evaluating the four point correlations and subsequently using the fact that the bath stays in thermal equilibrium (Eq.~\eqref{eq:bath_thermal}), we establish that the calculation of higher correlators can be decomposed into the evaluation of single-particle correlators, thus validating Wick's decomposition in the steady state,
\begin{equation}
\left<f_{a}^{k\dagger}f^{k\dagger}_b f_{c}^k f_{d}^k\right>=\left<f^{k\dagger}_{a}f^k_{d}\right>\left<f_{b}^{k\dagger}f_{c}^k\right>-\left<f^{k\dagger}_{a}f^k_{c}\right>\left<f_{b}^{k\dagger}f_{d}^k\right>.
\end{equation} 
 We therefore conclude that under such approximations of weak coupling and high frequency of the drive, the reduced density matrix of the system at asymptotic times is Gaussian and assumes a time-independent Gibbs form,
\begin{equation}
\tilde{\rho}(t)=\bigotimes_k {\cal N}_ke^{-\beta\sum\limits_{b}E_{b}^kf^{k\dagger}_{b}f^k_{b}}=\bigotimes_k\tilde{\rho}_k(t),
\end{equation}
where $\mathcal{N}_k$ normalizes the density matrix.
Reverting back to the actual frame of reference and observing at asymptotically large stroboscopic instants of time (t=$NT$), the final density matrix boils down to,
\begin{equation}
\rho_k(NT)\rightarrow F_k^{\dagger}(0)e^{-\beta\sum\limits_{b}E_{b}^kf^{k\dagger}_{b}f^k_{b}}F_k(0),
\end{equation}
where $F_k(t)$ is the unitary time-periodic kick operator (see Appendix.~\ref{Sec:solution}). Thus, the steady state stroboscopic density matrix is a Gibbs state in the Floquet Hamiltonian $H^F(k)$,
\begin{equation}\label{eq: density_matrix}
\rho(NT)=\bigotimes_k {\cal N}_ke^{-\beta\sum\limits_{\alpha,\beta}{a^{k\dagger}_{\alpha}H^F_{\alpha\beta}(k)a^k_{\beta}}},
\end{equation}
and decoupled for each $k\in{\rm BZ}$. As the stroboscopic steady state of the system is a Gibbs state with a temperature of that equal to the bath, its purity is completely determined by the reservoir temperature. As we establish in the following sections, the Chern topological classification can be extended to such states.

Further, the Gaussian nature explicitly implies that the stroboscopic steady state can be brought arbitrarily close to a pure state by reducing the temperature of the bath while preserving its topology.

\section{Macroscopic Electric Polarisation in the stroboscopic steady state}\label{sec:epol}
In this section, we recall the definition of the macroscopic electric polarisation of the bulk system in the asymptotic stroboscopic steady state. We establish that the macroscopic polarisation is in itself a many-body quantity which we evaluate for the stroboscopic steady state of the system. 
The stroboscopic steady state density matrix of the system can be written in terms of the real space creation and annihilation operators acting locally on each site of the Haldane model as,
\begin{equation}\label{eq:ss_rho}
\rho=\frac{e^{-\beta\sum\limits_{i,j}a^{\dagger}_iH^F_{ij}a_j}}{{\rm Tr}\left[e^{-\beta\sum\limits_{i,j}a^{\dagger}_iH^F_{ij}a_j}\right]},
\end{equation}
where $H^F$ is the Floquet Hamiltonian in real space and the index $i\equiv\{i,\alpha\}$ encompasses both the sublattice and the intra-sublattice index respectively. 

In this state, we evaluate the macroscopic electric polarisation vector of the system,
\begin{equation}
\vec{P}=P_1\vec{a}_1+P_2\vec{a}_2,
\end{equation}
which in the thermodynamic limit, reduces to \ct{vanderbilt09,resta94}
\begin{equation}\label{eq:pure_pol_1}
P_i=\sum\limits_{\alpha}{\rm Im}\int_{BZ}\braket{\psi_{k,\alpha}|\partial_{k_i}|\psi_{k,\alpha}}dk_{1}dk_{2},
\end{equation}
where $k$ denotes the single-particle momenta while $\alpha$ is the band indices.

We extend this definition of the polarisation vector for the mixed Gaussian steady state (Eq.~\eqref{eq:ss_rho}) as a weighted sum over the polarisation over all Floquet eigenstates weighted by their respective populations in the stroboscopic steady state, which reduces to (see Appendix.~\ref{sec:bulk_pol_ap}),
\begin{equation}\label{eq: pol_simple}
P_i={\rm Im}\sum\limits_{b}\ln\left[1+(L_i)_{bb}\right],
\end{equation}
such that components $(L_i)_{bb}$,
\begin{equation}\label{eq: pol_simple1}
(L_i)_{bb}=\prod\limits_{k}diag\{e^{-\beta E^k_{b}}\}e^{i\left(A_{i}^k\right)_{bb}\delta_i},
\end{equation}
where, $\delta_i=2\pi/L_i$, $L_i$ being the dimension of the system in the $i^{th}$ direction and  $E^k_{\alpha}$ are the Floquet quasi-energies and $\left(A_{i}^k\right)_{bb}$ is the $U(1)$ gauge connection over the band $b$ in the $i^{th}$ direction,
\begin{equation}
\left(A_{i}^k\right)_{bb}=\braket{\psi_{k,b}|\partial_{k_i}|\psi_{k,b}}.
\end{equation}

In the limit where the temperature of the bath goes to zero (i.e. $\beta\rightarrow\infty$), the many-body exponential weights (see Eq.~\eqref{eq:L_i} of Appendix.~\ref{sec:bulk_pol_ap}) predominantly selects the lowest energy band for each $k$ mode. In the $\beta\rightarrow\infty$ limit, the corresponding macroscopic polarisation approaches,
\begin{equation}
P_i\simeq {\rm Im}\ln(L_i)_{gg},
\end{equation}
where $(L_i)_{gg}$ is the product of the element of the matrix $L_i$ in the lowest quasi-energy state $\ket{g_k}$ over all $k$. Hence, the pure state polarisation reduces simply to,
\begin{equation}\label{eq:pure_pol}
P_1=\int_{k_{02}}^{k_{02}+1}\int_{k_{01}}^{k_{01}+1}dk_{1}dk_{2}\bra{g_k}\partial_{k_1}\ket{g_k}
\end{equation}
and like wise for $L_2$, where $k_0\equiv(k_{01},k_{02})$ is the origin of the Brillouin zone over which the integration is performed. The macroscopic polarisation is observed to express itself as a weighted sum of the polarisation over each band of the Floquet Hamiltonian. \\

However, it is well established that the macroscopic polarisation is not uniquely defined in a Chern non-trivial phase. Hence, despite its many-body nature, the macroscopic polarisation is not a measurable quantity in a Chern insulator. In the next section we utilise precisely this property of $P_i$ to extract out the topological information of the stroboscopic steady state and later propose a way to look at experimentally observable effects of the defined many-body polarisation.
\begin{figure}
	
	\begin{center}
		\includegraphics[width=0.48\textwidth,height=0.65\columnwidth]{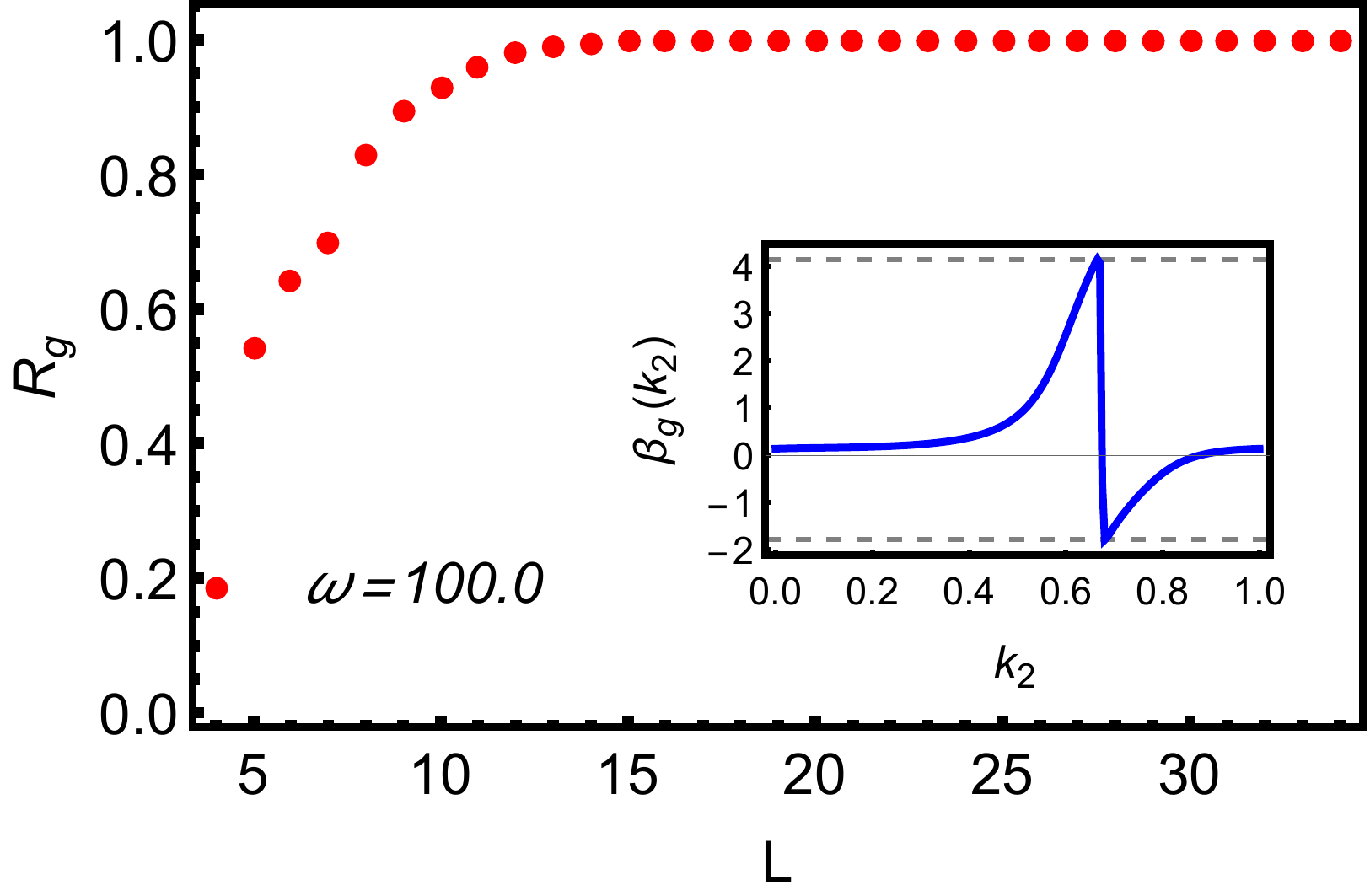}
	\end{center}

	\caption{(Color online) The fractional contribution of the Floquet ground state in the macroscopic polarisation $P_1$ with a finite temperature thermal state of the of the Floquet Hamiltonian.
		The bare Hamiltonian and the driving parameters are chosen to be $t_1=1.0$, $t_2=1.0$, $M=3\sqrt{3}t_2+0.05$, $\phi=-\frac{\pi}{2}$, $V_0=7.5$, $\beta=0.01$ in a $L\times L$ lattice.
		(Inset) The corresponding $\beta_g(k_2)$ function (defined in Eq.~\eqref{eq:beta}) of the Floquet ground state exhibiting a sharp branch change of $\Delta=-2\pi\times0.94$ and hence a corresponding Chern number $\mathcal{C}\simeq-1$ and thereby reflecting the topological character of the band. The parameters used are same as in (a) in a $200\times 200$ lattice with a periodic boundary condition.}
	\label{fig:5}
	
\end{figure}

\section{Chern number from the non-uniqueness of bulk electric polarisation}\label{sec:chern}
Following up the definition of the macroscopic bulk polarisation, in this section we further investigate the  anomaly that arises in the polarisation when the system is in a in a pure Chern non-trivial phase \ct{vanderbilt09}. Thereby it is shown that such a topological anomaly in the definition of the bulk polarisation does indeed survive in a mixed state. Let us recall that a  similar approach has already been successfully employed within a 
completely unitary set up in Ref. \ct{souvik201}.

Consider the evaluation of the integrals on the right hand side (RHS) of Eq.~\eqref{eq:pure_pol}. For every adiabatic shift in the centre of the Brillouin Zone, the electric polarisation vector changes proportionally to the Chern Invariant. If the system is in a pure state, for a shift in $\Delta k_0$ in the origin $k_0$ of the Brillouin zone,
\begin{equation}
P_i[k_0+\Delta k_0]-P_i[k_0]=\epsilon_{ij}(\Delta k_{0})_j{\cal C},
\end{equation}
where $\mathcal{C}$ is the Chern Number and $\epsilon_{ij}$ is the anti-symmetric symbol. We utilize this non-uniqueness of the electric polarisation to define the Chern number as,
\begin{equation}\label{eq:chern_diss}
\mathcal{C}= \epsilon_{ij}\frac{\Delta P_i[k_0]}{2\pi\Delta k_{0j}},
\end{equation}
where $P_i[k_0]$ is as defined in Eq.~\eqref{eq: pol_simple}. It is straight-forward to see from Eq.~\eqref{eq: pol_simple}-\eqref{eq:thermo} that when $\beta\rightarrow\infty$ or in the thermodynamic limit $L\rightarrow\infty$, $\mathcal{C}$ reduces to the Chern number of the lowest quasi-energy Floquet band, 
\begin{equation}\label{eq:pol_diff}
\begin{split}
\mathcal{C}\propto P_1[k_0+\Delta k_0]-P_1[k_0]=\\
-\Delta k_{02}\int_{k_{02}}^{k_{02}+1}dk_{2}\partial_{k_2}\beta_g(k_2),
\end{split}
\end{equation}
where $\mathcal{C}$ is the Chern number of the band $\ket{g_k}$ and,
\begin{equation}\label{eq:beta}
\beta_g(k_2)=-{\rm Im}\int_{k_{01}}^{k_{01}+1}dk_{1}\bra{g_k}\partial_{k_1}\ket{g_k}.
\end{equation}
The second equality in Eq.~\eqref{eq:pol_diff} states that the defined Chern number $\mathcal{C}$ counts the winding of the function $\beta_g(k_2)$ defined over the lowest quasi-energy state of the Floquet Hamiltonian. As shown in the inset of Fig.~\ref{fig:5}, owing to the topological nature of the state $\ket{g_k}$ and the Floquet Hamiltonian, the function $\beta_g(k_2)$ indeed shows a branch singularity with a jump $\Delta$ and thereby reflecting a non-zero Chern number,
\begin{equation}
\mathcal{C}=\frac{\Delta}{2\pi}.
\end{equation}
This proves that the stroboscopic steady state is indeed a topologically non-trivial thermal state. Furthermore, it has a non-trivial Chern number at any finite temperature in the thermodynamic limit. Therefore, it is also possible to engineer a stroboscopic pure state for $\beta \to \infty$ having non-trivial topology in asymptotic time.
 This is a consequence of equilibration of the stroboscopic system in the Floquet ground state when the temperature of the bath approaches zero (see Eq.~\eqref{eq: density_matrix}). 

{At point, let us address the question whether the zero-temperature topology survives}  for a finite temperature steady state which is   in equilibrium with  the Floquet Hamiltonian when  the polarisation has the many body form 
shown in Eqs. \eqref{eq: pol_simple} and {hence, is a mixed state.}

However, even at a finite temperature, the polarisation in a Gibbs' state reduces to that of the lower purity band in the thermodynamic limit. This can be appreciated through Eq.~\eqref{eq: pol_simple1} as,
\begin{equation}\label{eq:thermo}
(L_i)_{bb}=\left(e^{-\beta L^2\sum\limits_{k}\frac{1}{L^2}E^k_{b}}\right)\prod\limits_{k}e^{i\left(A_{i}^k\right)_{bb}\delta_i},
\end{equation}
for a $L\times L$ lattice. In the thermodynamic limit ($L\rightarrow\infty$), the exponential weights the quantity over just the lowest quasi-energy band. We demonstrate this with the stroboscopic steady state in Fig.~\ref{fig:5} where the fraction,
\begin{equation}\label{eq:pol_fraction}
R=\frac{P_g}{P_g+P_e},
\end{equation}
such that $P_g$ and $P_e$ are the polarisations i.e., $P_1$ evaluated over the Floquet ground state and excited state respectively.

To elaborate, in the thermodynamic limit, the exponential weight factors due to each band in Eq.~\eqref {eq: pol_simple1}, {\it projects out} the contribution of the lowest quasi-energy band in the expression for the macroscopic polarisation while exponentially suppressing the contribution of the other bands. It thus follows that the topological invariant defined in Eq.~\eqref{eq:chern_diss} reflects the topology of the ground state of the Floquet Hamiltonian at all finite temperatures. This can be intuitively understood as the continuous deformation of a pure state into a mixed state density matrix is a smooth transformation as long as the purity gap does not close and the topology of a state is not expected to change under any continuous deformation. 
For a generic Gaussian density matrix with a closed purity gap, the many-body weights,
\begin{equation}
W_b=e^{-\beta L^2\sum\limits_{k}\frac{1}{L^2}p^k_{b}},
\end{equation}
become degenerate for two or more purity band with eigenvalues $p^k_b$. This prevents the statistical projection of the polarisation to that of the lowest purity band. Hence, the defined Chern number no longer reflects a $U(1)$ curvature, in the process destroying its quantization. However, as long as the Gibb's state have non-degenerate purity bands, the Chern number remains perfectly quantised and can only change while crossing a purity band inversion which in the present case, is nothing but the Floquet bands.

This has already been established {in the context of mixed state dynamical 
quantum phase transitions  characterised by the interferometric phase where the dynamical analogue of the partition function is defined as,
\begin{equation}
Z={\rm Tr}\left[e^{-iHt}\rho(0)\right]=\braket{U(t)}_0,
\end{equation}
for an initial state $\rho_0$ undergoing time evolution generated by the final Hamiltonian $H$ following a quench. Here the role of the spatial translation operator in the definition of the topological invariant} is played by the temporal translation operator $U(t)$, i.e., the propagator itself \ct{utso172,budich17,souvik18}. This is tantamount  to saying that the topology of finite temperature Gibbs state  can only be  altered by changing the temperature without crossing the infinite temperature point. Such a transition is therefore highly unlikely in a thermodynamically large many-body quantum system.\\

Invoking upon the definition of Chern number as in Sec.~\ref{sec:chern}, we note that the out of equilibrium Chern number $\mathcal{C}(t)$ reduces to nothing but the Chern number of the lowest purity band of the density matrix in the thermodynamic limit although for a Gaussian state, the generalisation to incorporate higher order correlators is necessary to classify non-Gaussian mixed states. Similar arguments hold when the periodic perturbation is subjected to a perfectly adiabatic temporal variation. In such situations, the stroboscopic system is continuously in equilibrium with the instantaneous Floquet hamiltonian. It thus contains the complete information of the topology of the Floquet Hamiltonian in its lowest purity band. 


\section{Particle current in the stroboscopic steady state}\label{sec:current}
In Sec.~\ref{sec:pop}, we argued that the stroboscopic steady state is a Gaussian density matrix. We explicitly verify this in this section by evaluating the total particle current flowing between the system and the bath in the stroboscopic steady state. By establishing that the mean number of particles in the system is indeed stationary in the stroboscopic steady state, we discuss  the stroboscopic bulk-boundary correspondence when the steady state is pure.

\subsection{System-reservoir particle current in the steady state}\label{sec:current1}

The Heisenberg evolution of the system and bath operators can be explicitly written down in the real space as,
\begin{equation}
\begin{split}
i\partial_t A_{ni}=\epsilon_{i}A_{ni}(t)+\sum\limits_{\alpha}\lambda_{i\alpha}a_{n\alpha}(t),\\
i\partial_t a_{n\alpha}=\sum\limits_{\beta,m} H_{\alpha,\beta,n,m}(t)a_{\beta m}(t)+\sum_{\mu}\lambda^{*}_{\mu\alpha}A_{n\mu}(t).
\end{split}
\end{equation}
Solving these set of coupled dynamical equations for the time-dependent system and bath degrees of freedom enables one to identify the asymptotic steady state in a finite size system. For asymptotically large times under the approximation of weak coupling and high frequency, we obtain,
\begin{equation}\label{eq:Ani}
A_{ni}(t)=\sum\limits_{\alpha}\int_{-\infty}^{\infty}d\Delta~e^{-i\Delta t}\frac{\lambda_{i\alpha}\tilde{a}_{n\alpha}(\Delta)}{\Delta-\epsilon_i},
\end{equation}
where,
\begin{eqnarray}
\tilde{a}_{m\alpha}(\Delta)&=&\int\limits_{0}^{\infty}a_{m\alpha}(t)e^{i\Delta t}dt \nonumber\\
&=&\sum\limits_{m,\beta,\gamma,i,n\in{\rm Z}}\frac{Y^{(n)\dagger}_{\beta\gamma}\lambda^{*}_{i\gamma}I_{m\alpha,\beta}^i(\Delta)A_{mi}(0)}{\epsilon_i-E^{(n)}_{\beta}+i\tilde{\Pi}^{\prime}_{\beta}(n\omega-\epsilon_i)}I^i(\Delta)\nonumber\\
&=&\int\limits_{0}^{\infty}Y(t)e^{-i(\epsilon_i-n\omega-\Delta)}dt,
\end{eqnarray}
with $Y^{(n)}$ and $\tilde{\Pi}$ being the corresponding real space quantities similar to as defined in Eq.~\eqref{eq:Y} and Eq.~\eqref{eq:Pi}.
Under the approximations in Sec.~\ref{sec:rot} and Sec.~\ref{sec:pop}, evaluating the equation in  Eq.~\eqref{eq:Ani}, we arrive at,
\begin{equation}\label{eq:Ainf}
A_{nj}(t)=A_{nj}(0)e^{-i\epsilon_j t}\sum\limits_{\alpha,m\in{\rm Z}}\frac{\tilde{\Pi}_{\alpha\alpha}^{mm}(\epsilon_j)}{\epsilon_j-E^{(m)}_{\alpha}+i\tilde{\Pi}^{\prime}_{\alpha}(m\omega-\epsilon_j)},
\end{equation}
and
\begin{equation}\label{eq:ainf}
a_{m\alpha}(t)=\sum\limits_{l,\beta,\gamma,n,i}\frac{Y^{(n)}_{m\alpha,\beta}Y^{(n)\dagger}_{\beta\gamma}\lambda_{i\gamma}^{*}e^{-i\epsilon_i t}}{\epsilon_i-E_{\beta}^{(n)}+i\tilde{\Pi}^{\prime}_{\beta}(n\omega-\epsilon_i)}A_{li}(0).
\end{equation}
We define the total particle current flowing between the reservoir and the system at each site as,
\begin{equation}\label{eq:Jsb}
\braket{J_{SB}}=\sum\limits_{n\alpha}\braket{J_{n\alpha}}=\sum\limits_{in\alpha}\left[\lambda_{n\alpha}\braket{A^{\dagger}_{ni}a_{n\alpha}}-cc\right],
\end{equation}
where the expectation is taken over the steady state distributions and $cc$ is the complex conjugate. Substituting Eq.~\eqref{eq:Ainf} and Eq.~\eqref{eq:ainf} in the expression for system-math current Eq.~\eqref{eq:Jsb}, we establish that the total system-bath current vanishes in the steady state,
\begin{equation}
\begin{split}
\braket{J_{SB}}=\sum\limits_{n,\delta,i,m}\frac{|\tilde{\Pi}^{nn}_{\delta\delta}(\epsilon_i)|^2}{|\epsilon_i-E_{\beta}^{(n)}+i\tilde{\Pi}^{\prime}_{\beta}(n\omega-\epsilon_i)|^2}\braket{A^{\dagger}_{mi}A_{mi}}\\
-cc=0.
\end{split}
\end{equation}
Thus, the mean particle number in the system becomes stationary in the stroboscopic steady state, implying that the steady state stroboscopic density matrix of the system is completely determined by the single-particle correlations. 

\subsection{Particle current in the steady state stroboscopic system }
Although in Sec.~\ref{sec:chern}, we establish that the steady stroboscopic state can indeed be topologically {non-trivial}, the manifestation of this topology through a bulk-boundary correspondence is what we address in this section. However, we investigate the existence of such a correspondence only if the steady stroboscopic state is pure, i.e., the bath is maintained at near zero absolute temperatures.\\

It is established that in a topologically non-trivial phase, the Haldane model hosts robust chiral curents localized at the edges of a finite size system.
The boundary currents under semi-open boundary conditions in the steady state stroboscopic density matrix of the system as in Eq.~\eqref{eq:ss_rho} must therefore be localised on the edge when the bath is taken to be at a very low temperature. In the $\beta\rightarrow 0$ limit, the stroboscopic current reduces perfectly to be that over the ground state of the Floquet Hamiltonian $H^F(k)$, which having a non-zero Chern number, is topologically non-trivial. The Chern number $\mathcal C$ being defined as an anomaly in the macroscopic electric polarisation of the system (Eq.\eqref{eq:chern_diss}), it is naturally expected to observe a stroboscopic edge-localised chiral current flowing in the steady state system \ct{xiao10}.
Although even in a finite temperature stroboscopic state, the topological invariant reduces to that of the ground state of the Floquet Hamiltonian, the population does not. That is, all the bands of the Floquet Hamiltonian are partially occupied in a photon sector and therefore, a generic bulk-boundary correspondence is not expected \ct{xiao10,rivas13}.\\

  Although topology of the pure steady state is expressed as a stroboscopic bulk boundary correspondence, the existence of such an observable phenomena is an area of further investigation. However, in the next section we argue that there exists other many body observable phenomena which might exhibit the Chern non-triviality of the stroboscopic steady state even when it is mixed.\\

\section{Conclusion and experimental connections}
\label{sec_conclusion}
Although there has been a significant development  in the dynamical engineering of Chern insulating Hamiltonians mostly under periodic driving protocol, the topology of the non-equilibrium state of a Chern insulator and corresponding physical observables, especially in
a dissipative environment, is still a developing area of research. We address this issue by redefining the topological invariant in out-of-equilibrium systems such that it incorporates the many-body filling of the density matrix purity bands. Considering a topological Haldane model, we probe the steady state behaviour of the many- body Chern invariant when the system is driven periodically in a  dissipative ambience. 
 
 In this non-unitary set up, to surmount the competition between locality and topological non-triviallity, posed by the Lindblad master equation approach \ct{bardyn13,goldstein19},  a micro-structured free fermionic bath is considered which acts quasi-locally and independently on each unit cell of the Haldane model and  an equation of motion approach is implemented. Although the system bath coupling is allowed to be different at each sublattice, each unit cell is assumed to couple independently and uniformly to the bath. This ensures that the translational invariance of the lattice remain intact even for a finite non-zero coupling with the  bath. The bath is further chosen to be in thermal equilibrium with  its energy modes occupied according to a Fermi distribution function at a finite temperature.\\

 Starting from a topologically trivial equilibrium state, the Haldane model is subjected to a time-periodic drive while simultaneously maintaining the coupling to the bath. The amplitude and frequency of the drive is so chosen that the resulting Floquet Hamiltonian is topologically non-trivial.

 Under the action of both the external drive and the bath, the complete system evolves dynamically as a coupled system from which we eliminate the degrees of freedom of the bath. The resulting degrees of freedom of the reduced system then follow a non-unitary dynamics while being simultaneously subjected to a  periodic driving. Under the approximations of weak system bath coupling and a sufficiently high frequency of the drive, we have  established that the stroboscopic steady state of the system is a finite temperature Gibbs state in equilibrium with the Floquet Hamiltonian. To further corroborate  this observation, we evaluate the total system bath particle flow in the stroboscopic steady state and establish that the mean particle number is indeed conserved in the steady state. Also, the higher fermionic correlations in the stroboscopic steady state exhibit Wick's decomposition into single particle correlation functions. We establish that the stroboscopically observed system relaxes to a steady state precisely at the temperature maintained for the bath.\\
 
 Further, to topologically classify the stroboscopic steady state, the many body bulk electric polarisation of the system has been utilized. We recall that due to the lack of well-localized Wannier functions, the macroscopic polarisation is not a well defined quantity in a Chern  insulator which is in a non-trivial phase. This is reflected in the fact that in the topological phase, the electric polarisation vector is not uniquely defined for a Chern Insulator. Using precisely this non-uniqueness of the bulk electric polarisation, we distinguish between topologically trivial and non-trivial steady states. The many body nature of the defined Chern invariant is evident in the incorporation of purity dependent weights representing the contribution of all the purity bands of the steady state stroboscopic density matrix in the Chern invariant. Nevertheless, in the thermodynamic limit, the redefined Chern invariant reduced to just the Chern number of the lowest purity band or in this case to that of the lowest quasi-energy eigenstate of the Floquet Hamiltonian. Also, in the pure state limit, the defined Chern number reduces to the unitary quantity of the sum of berry phases evaluated over all completely filled single particle states. We further establish that by introducing a cut-off in the coupling energy of the system-bath interaction, a differentially controlled occupation of different photon sectors can be obtained in the stroboscopic steady state. Further, if the driving frequency is much larger than the cut-off energy scale, it has been established that only the zero photon sector of the Floquet Hamiltonian can be made to be  dominantly occupied in the steady state and hence, results in a truely topological Floquet steady state.\\
 
  To probe a possible stroboscopic bulk boundary correspondence in the case of a pure topological steady state, we examine the inter-system single-particle current in the stroboscopic steady state. The steady state density matrix being thermal, we argue that the single particle currents flowing within the system shall mimic that of a system in an equilibrium topological phase for a low-temperature bath.
 
 To summarise, the Floquet Hamiltonian in the completely unitary situation being topologically non-trivial, we infer that it is indeed possible to engineer topologically non-trivial Chern states at any finite temperature with such a micro-structured bath. The bath is presumed to act as a substrate to the 2D Haldane lattice to which the system dumps excess energy absorbed from the drive 
 and thermalises to a stroboscopic Gaussian state.

 The bath considered in the paper can be experimentally realised through detailed construction of a substrate that preserves the basic symmetries of graphene at least for a finite number of unit cells. The periodic driving protocol introduced can be experimentally realised through the generation of pseudo-magnetic fields \cite{suman19} in graphene \cite{levy10}. Recently, the Floquet anomalous Hall current have been probed in transport experiments \cite{mcIver20}. Although finite Hall currents are observed in the Floquet topological phases, a dominant contribution to the current is seen to have arisen out of a non-topological population imbalance of photocarriers in the Brillouin Zone. This has also been numerically verified through phenomenological microscopic models harnessing a master equation approach to incorporate dissipation \cite{sato19,sato19prb}. We propose a robust method to engineer thermal Floquet topological phases with controlled temperatures while at the same time providing a handle on the asymptotic filling of Floquet bands, and are therefore expected to exhibit topological currents in similar transport experiments.

 The generalised Chern number as described in the stroboscopic steady state of the driven dissipative Haldane system, being dependent on the steady state correlations is expected to be measurable at all temperatures through many-body observables. Such an experimental set up has already been suggested \ct{diehl18} in the case of finite temperature topological phase transitions in 1D many body systems.\\
 In Mach-Zender type of interferometric set-ups with particular TEM modes of light, the macroscopic polarisation of the system manifests as a phase shift as light passes through the system \ct{diehl18}. In a Chern insulator, the macroscopic polarisation is not in itself a topological quantity like in a 1D situation. However, the polarisation vector shows a change proportional to the Chern number of the system under an adiabatic parallel shift of the reciprocal lattice vectors. 
 Since the driven dissipative system thermalizes with the Floquet Hamiltonian. The adiabatic generation of a synthetic $U(1)$ gauge field in the Floquet Hamiltonian \ct{goldman14,anatoli16} will precisely serve this purpose as the quasi-momentum couples to the synthetic gauge potential. This in turn is expected to manifest as a topological phase shift of light in an interferometric set-up. Thus, with a careful design of the substrate, the generalised finite temperature topological Chern phases of Floquet systems can be realised in state of the art experiments.\\

\begin{acknowledgments}
	We acknowledge  Arijit Kundu, Anatoli Polkovnikov and Diptiman Sen  for helpful discussions and critical comments. We acknowledge Sourav Bhattacharjee and Somnath Maity for  discussions. SB acknowledges PMRF, MHRD India for financial assistance. AD acknowledges financial support from SPARC program, MHRD, India. We also acknowledge ICTS, Bangalore, India where some part of the work was done.
\end{acknowledgments}

\appendix

\section{A brief review on Haldane model of graphene :}\label{appendix:C}

Interestingly, the Haldane model with explicitly broken time reversal symmetry is known to host topologically non-trivial phases for certain parameter regions. The topology of the Hamiltonian is esentially the homotopy classification of the map $(k_1,k_2)\rightarrow H^k(k_1,k_2)$ in reciprocal space and is characterized by the gauge invariant Chern topological invariant,
\begin{equation}
C=\frac{1}{\left(2\pi\right)^2}\int_{BZ}dk_1dk_2\mathcal{F}_{12}(\ket{\psi_k}),
\end{equation}
where, $\mathcal{F}_{12}(\ket{\psi_k})$ is the $U(1)$ curvature defined over the ground state $\ket{\psi_k}$ of the Hamiltonian $H^k$, i.e.,
\begin{equation}
\begin{split}
\mathcal{F}_{12}(\ket{\psi_k})= \partial_{k_2}\langle{\psi_k|\partial_{k_1}|\psi_k}\rangle-\partial_{k_1}\langle{\psi_k|\partial_{k_2}|\psi_k\rangle}.
\end{split}
\end{equation}

The Chern invariant is integer quantized as long as the Hamiltonian $H^k$ does not approach a QCP where the Chern number becomes ill-defined. Different integer values of the Chern number characterize distinct topological phases separated by QCPs.

{Each point on the Bravias lattice can be referenced in terms of the Bravias lattice vectors, i.e.,
	\begin{equation}
	\vec{a}=n_1\vec{a}_1+n_2\vec{a}_2,
	\end{equation}
	where the vectors $\vec{a}_1$ and $\vec{a}_2$ span the Bravias lattice and $n_1,n_2$ are integers.
	We choose the vectors $\vec{a}_1$ and $\vec{a}_2$ to be the next nearest neighbour hopping vectors such that,
	\begin{equation}\label{eq:lattice_supp}
	\begin{split}
	\vec{a}_1=\vec{\Delta}_{22},\\
	\vec{a}_2=-\vec{\Delta}_{21},
	\end{split}
	\end{equation}
	
	\begin{figure*}
		\begin{subfigure}{0.45\textwidth}
			\includegraphics[width=\textwidth]{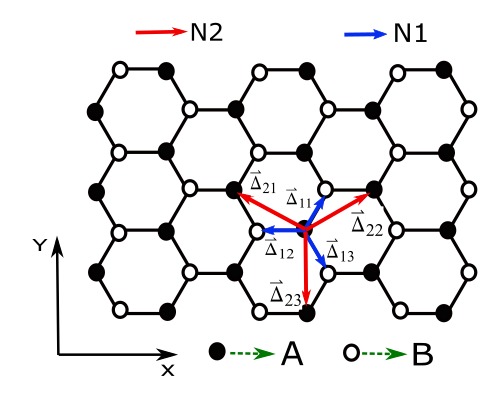}
			\caption{} \label{fig_1a_sup} 
		\end{subfigure}
		\quad\quad\begin{subfigure}{0.45\textwidth}
			\includegraphics[width=\textwidth]{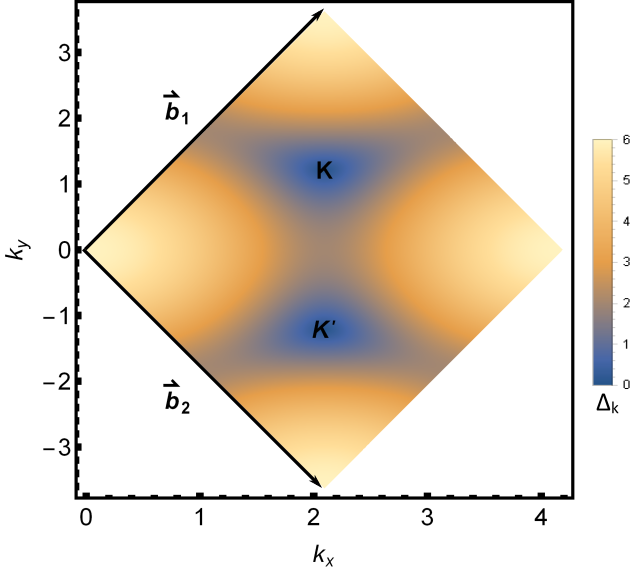}
			\caption{}	\label{fig_1b_sup}
		\end{subfigure}
		\caption{(Color online) (a) The hexagonal graphene lattice showing the nearest neighbour (N1) and next-nearest neighbour (N2) hopping vectors $\vec{\Delta}_{1i}$ and $\vec{\Delta}_{2i}$, respectively, where the lattice constant is set to be $a=1$. The hollow and the filled atoms represent the B and A sublattices respectively. (b) The Brillouin zone of graphene spanned by the reciprocal lattice vectors $\vec{b}_1$ and $\vec{b}_2$ containing two inequivalent Dirac points $K$ and $K^{\prime}$ (the cartesian directions has been labelled by $k_x$ and $k_y$ respectively). The color density shows the absolute value of the bandgap 
			$\Delta_k$ of the reciprocal space graphene Hamiltonian showing vanishing gaps at the Dirac points for a $600\times 600$ lattice size having the N1 hopping strength $t_1=1.0$ and the N2 hopping $t_2=0$.}
	\end{figure*}

	where $\vec{\Delta}_{2i}$ are the $N2$ vectors as shown in Fig.~\ref{fig_1a_sup}.
	
	Invoking the discrete translational invariance of the Hamiltonian one can employ a discrete Fourier transform to decouple the Hamiltonian $H(t)$ in momentum space. The reciprocal space is spanned by the reciprocal lattice vectors $\vec{b}_1$ and $\vec{b}_2$, i.e. every reciprocal lattice point can be represented as,
	\begin{equation}
	\vec{b}=k_1\vec{b}_1+k_2\vec{b}_2,
	\end{equation}
	where, $k_1$, $k_2\in[0,1)$. We choose a rhomboidal Brillouin zone spanned by reciprocal lattice vectors $\vec{b}_1$ and $\vec{b}_2$ (see Fig.~\ref{fig_1b_sup}) containing two independent Dirac points $K$ and $K^{\prime}$}. In our choice of representation,
\begin{equation}
\vec{b}_1=\frac{2\pi}{3a}\{1,\sqrt{3}\}~~\text{and}~~\vec{b}_2=\frac{2\pi}{3a}\{1,-\sqrt{3}\},
\end{equation}
where we have chosen $a=1$. The corresponding inequivalent Dirac points in the Brillouin zone shown in Fig.~\ref{fig_1b_sup} are given by,
\begin{equation}
K=\frac{2\pi}{3}\left(1,\frac{1}{\sqrt{3}}\right)~~\text{and}~~K^{\prime}=\frac{2\pi}{3}\left(1,-\frac{1}{\sqrt{3}}\right).
\end{equation}

The bare Haldane Hamiltonian gets decoupled in the momentum space where $H^0(k)$ can be written in the basis $\ket{k,A}$ and $\ket{k,B}$ as,
\begin{equation}
H^0(k)=\vec{h}(k).\vec{\sigma} =h_x(k) \sigma_x+h_y(k) \sigma_y+h_z(k) \sigma_z,
\label{eq_hamil_k}
\end{equation}
such that,
\begin{equation}\label{eq:bloch_ham}
\begin{split}
h_x(k)=-t_1\sum\limits_{i=1}^{3}\cos{\left(\vec{k}.\vec{\Delta}_{1i}\right)},\\
h_y(k)=-t_1\sum\limits_{i=1}^{3}\sin{\left(\vec{k}.\vec{\Delta}_{1i}\right)},\\
h_z(k)=M-t_2\sin{\phi}\sum\limits_{i=1}^{3}\sin{\left(\vec{k}.\vec{\Delta}_{2i}\right)},
\end{split}
\end{equation}
$\vec{\Delta}_{1i}$ and $\vec{\Delta}_{2i}$ are the nearest neighbour and next nearest neighbour lattice vectors respectively (see Fig.~\ref{fig_1a_sup}) chosen to be,

\begin{widetext}
\begin{equation}
\begin{split}
\vec{\Delta}_{11}=\frac{a}{2}\{1,\sqrt{3}\},~~\vec{\Delta}_{12}=\{-a,0\},~~\vec{\Delta}_{13}=\frac{a}{2}\{1,-\sqrt{3}\} ~~\text{and},\\
\vec{\Delta}_{21}=\frac{a}{2}\{-3,\sqrt{3}\},~~\vec{\Delta}_{22}=\frac{a}{2}\{3,\sqrt{3}\},~~\vec{\Delta}_{23}=\{0,-a\sqrt{3}\},
\end{split}
\end{equation}
\end{widetext}
in the cartesian frame (Fig.~\ref{fig_1a_sup}) where we have chosen the lattice parameter $a=1$.

\section{Solving for the reduced system}\label{Sec:solution}

The dynamics of the complete system comprising of the Haldane model and the resercoir, is dictated by two coupled differential equations as in Eq.~\eqref{eq:eom}. Simultaneously solving the above set of equations and eliminating the bath degrees of freedom we obtain one single equation which the governs the reduced system comprising of just the Haldane model as,
	\begin{equation}\label{eq:red_eqn}
	\begin{split}
	i\frac{da_{\alpha}^k}{dt}=\sum_{\beta} H_{\alpha,\beta}^k(t)a_{\beta}^k(t)\\
	-i\int_0^t\sum_{\eta}\Pi_{\alpha,\eta}^k(t^\prime) a_{\eta}^k(t-t^{\prime})dt^{\prime}+\zeta_{\alpha}^k(t),
	\end{split}
	\end{equation}
	where,
	\begin{equation}
	\begin{split}
	\Pi_{\alpha,\beta}^k(t)=\sum_{\mu}\lambda^*_{\mu,\alpha}\lambda_{\mu,\beta}e^{-i\epsilon^k_{\mu}t}~~\text{and,}\\
	\zeta_{\alpha}^k(t)=\sum_{\nu}\lambda_{\nu,\alpha}e^{-i\epsilon^k_{\nu}t}A_{\nu}(0),
	\end{split}
	\end{equation}
Such that the total scattering amplitude after time $t$,
\begin{equation}
\Lambda^k_{\alpha,\beta}(t)=\int_0^{t}\Pi_{\alpha,\beta}(t^{\prime})dt^{\prime}.
\end{equation}
	Further, employing transformation to a rotating frame of reference generated by a {\it time periodic} unitary transformation,  we obtain an effective Hamiltonian with no explicit time-dependence,
	\begin{equation}
	H^{\rm eff}_k=F_k^{\dagger}(t)H^k(t)F_k(t)-iF_k^{\dagger}(t)\partial_tF_k(t).
	\end{equation}\\

	Denoting the creation and destruction operators of the eigenmodes of the effective Hamiltonian by $f_{\alpha}^k(t)$ and $f_{\alpha}^{k\dagger}(t)$, we note recall that they differ from operators $a_{\alpha}^k(t)$ as,
	\begin{equation}
	f_{b}^k(t)=\sum_{\beta}Y_{b,\beta}^{k\dagger}(t)a_{\beta}^k(t),
	\end{equation}
	where the index $b$ refers to ehe effective Hamiltonian bands and, 
	\begin{equation}\label{eq:Y}
	Y^k=F^k(t)V^k
	\end{equation}
	with $V^k$ being the unitary operator that diagonalises the effective Hamiltonian $H^{\rm eff}_k$. Under the rotating wave and the weak coupling approximation as discussed in Sec.~\ref{sec:rot}, dynamical equation for the effective mode simplifies to include only the diadonal self-energy terms in asymptotically large times,
	\begin{equation}\label{eq:diagonal Floquet_a}
	\begin{split}
	i\partial_t f_b^k=E^k_b f_b^k(t)-i\sum_{n}\int_{0}^t\tilde{\Pi}^{k,nn}_{b,b}f_{b}^k(t-t^\prime)e^{in\omega t^\prime}dt^\prime\\
	+i\sum_{b^{\prime}}Y^{k\dagger}_{b,b^{\prime}}\zeta_{b^{\prime}}^k(t).
	\end{split}
	\end{equation}
  The equation Eq.~\eqref{eq:diagonal Floquet_a} can be solved using a Fourier transform to obtain,
	\begin{equation}\label{eq:f}
	f_{b}^k(t)=\sum_{i,b^{\prime},n}\frac{Y^{k(n)\dagger}_{b,b^{\prime}}\lambda^*_{i,b^{\prime}}e^{-i(\epsilon^k_i-n\omega)t}}{\epsilon^k_i-E^{k(n)}_b+i\tilde{\Pi}^{\prime k}_\alpha(n\omega-E^k_b)}A^k_{i},
	\end{equation}
	where, 
	\begin{equation}
	\begin{split}
	\tilde{\Pi}^{\prime k}_b(x)=\sum_{n}\int_0^{\infty}dt^{\prime} \tilde{\Pi}^{nn}_{bb}(t^\prime)e^{i(x-n\omega)t^{\prime}}~~\text{and}~~\\
	E_{b}^{k(n)}=E_{b}^{k}+n\omega
	\end{split}
	\end{equation}
	Observed at stroboscopic intervals $f_{b}^k(NT)$ gives the stroboscopically evolved Heisenberg Floquet annihilation operator.
	Using Eq.~\eqref{eq:f}, it is straight forward to obtain the expectation of occupation into the Floquet states at asymptotically long stroboscopic time,
	\begin{equation}\label{flo_exp}
	\left<f_{b}^{k\dagger}f_{b}^k\right>=\sum_{n}\int_{-\infty}^{\infty}d\epsilon^k_p\frac{W_{bb}^{k nn}(\epsilon^k_p)\left<A^{k\dagger}_{b}A^k_{b}\right>}{\left|\epsilon^k_p-E^{k(n)}_{b}+i\tilde{\Pi}^{\prime k}_b(n\omega-E^k_b)\right|^2},
	\end{equation}
	where,
	\begin{equation}
	\begin{split}
	W^{knn}_{bb}=\left[Y^{k(n)\dagger}W^kY^{k(m)}\right]_{bb},\\
	W^{k}_{\alpha\beta}=\int_0^{\infty}dt\Pi^k_{\alpha\beta}(t)e^{i\Omega t}.
	\end{split}
	\end{equation}
The fourrier transforms $Y^{k(n)}$ of the unitary matrix $Y^k$ are dependent on the frequency of external periodic drive in the high frequency approximation as, 

\begin{equation}
Y^{k(n)}\sim{\it O}\left(\left(\frac{A^2}{\omega}\right)^n\right),
\end{equation}
where, $A$ is the amplitude of the time-periodic drive.

\section{Evaluation of the bulk polarisation}\label{sec:bulk_pol_ap}

{Given the asymptotic steady state, it is straight forward to evaluate the macroscopic polarisation in the quasi-momentum picture.The steady state being Gaussian, the expectation value of all stroboscopic quadratic observables in the steady state reduces to the expectation over an effective action which is  Gaussian in grassmannian fields and are determined solely by the Fermi distribution function,} (Eq.~\eqref{eq:ss_rho}),
	\begin{equation}
	\left<\hat{T}_i\right>=\det\left[\mathbb{ I}-f_{FD}(H^F)+f_{FD}(H^F)T_i\right],
	\end{equation}
	where,
	\begin{equation}
	f_{FD}(H^F)=\frac{\mathbb{I}}{e^{-\beta H^F}+\mathbb{I}}~~\text{and}~~(T_i)_{l,m}=\delta_{lm}e^{i\delta_i x_i^l}.
	\end{equation}
	Hence, the polarisation in the $i^{th}$ direction can be compactly written as,
	\begin{equation}\label{eq:avgP}
	P_i={\rm Im}\ln\det\left[\mathbb{ I}-f_{FD}(H^F)+f_{FD}(H^F)T_i\right].
	\end{equation}
	The action of the unitary translation matrices $T_i$ in the momentum space is understood to be,
	\begin{equation}\label{eq:Tk}
	\begin{split}
	T_1=\sum\limits_{k}\ket{k_1+1,k_2}\bra{k_1,k_2}~~\text{and}~~T_2\\
	=\sum\limits_{k}\ket{k_1,k_2+1}\bra{k_1,k_2}
	\end{split}
	\end{equation}
	and the Floquet Hamiltonian is expressed as,
	\begin{equation}
	\begin{split}
	H^F=\bigotimes_kH^F(k)\equiv\sum\limits_{k}H^F(k)\ket{k}\bra{k}=\\
	\sum\limits_{k}D_k diag\{E^k_b\}D^{\dagger}_k\ket{k}\bra{k},
	\end{split}
	\end{equation}
	where the unitary operators $D_k$ diagonalise the Floquet Hamiltonian for each $k$ mode. In the momentum space the operators $f_{FD}(H^F)$ therefore assumes the form,
	\begin{equation}\label{eq:fk}
	f_{FD}(H^F)=\sum\limits_{k}D_k diag\{f_{FD}(E^k_b)\}D^{\dagger}_k\ket{k}\bra{k}.
	\end{equation}
	Where by '$diag\{E^k\}$' we signify the diagonal matrix with the quasi-energies as its diagonal elements.
	Using Eq.~\eqref{eq:Tk} and Eq.~\eqref{eq:fk} in Eq.~\eqref{eq:avgP}, after a few algebraic simplifications we obtain the expectation of the polarisation on the $i^{th}$ direction as,
	\begin{equation}\label{eq:pol}
	\left<P_i\right>={\rm Im}\ln{\det}^{\prime}\left[\mathbb{I}+L_i\right],
	\end{equation}
	where {we have explicitely evaluated the determinant over the diagonal momenta blocks} and the determinant ${\det}^{\prime}$ now acts only on the 'band-space', i.e., over the sub-lattice degrees of freedom for a particular $k$-mode, and $L_i$ is a matrix in the sublattice-basis,{
		\begin{equation}\label{eq:L_i}
		\begin{split}
		L_1=\prod\limits_{k}diag\left[\frac{f_{FD}(E^k_b)}{1-f_{FD}(E^k_b)}\right]\left(D^{\dagger}\right)_{k_1+1,k_2}\left(D\right)_{k_1,k_2}=\\
		\prod\limits_{k}diag\{e^{-\beta E^k_b}\}\left(D^{\dagger}\right)_{k_1+1,k_2}\left(D\right)_{k_1,k_2},~~\text{and}~~\\
		L_y=\prod\limits_{k}diag\left[\frac{f_{FD}(E^k_b)}{1-f_{FD}(E^k_b)}\right]\left(D^{\dagger}\right)_{k_1,k_2+1}\left(D\right)_{k_1,k_2}=\\
		\prod\limits_{k}diag\{e^{-\beta E^k_b}\}\left(D^{\dagger}\right)_{k_1,k_2+1}\left(D\right)_{k_1,k_2}.
		\end{split}
		\end{equation}}\\

	In the continumm limit ($\delta_i\ll 1$), 
	\begin{equation}
	\begin{split}
	\left[\left(D^{\dagger}\right)_{k_1+1,k_2}\left(D\right)_{k_1,k_2}\right]_{bb^{\prime}}=\\
	\braket{\psi_{k_1+1,k_2,b}|\psi_{k_1,k_2,b^{\prime}}}\simeq e^{i\left(A_{1}^k\right)_{bb^{\prime}}\delta_1},
	\end{split}
	\end{equation}
	where $\ket{\psi_{k,b}}$'s are the eigenvectors of the Floquet Hamiltonian $H^F(k)$ and $\left(A_{1}^k\right)_{bb^{\prime}}=\braket{\psi_{k,b}|\partial_{k_1}|\psi_{k,b^{\prime}}}$.

\end{document}